\documentclass[longauth]{aa}
\usepackage{amsmath,amssymb,amsfonts}
\usepackage{graphicx}
\usepackage{txfonts}
\usepackage{enumerate} 
\usepackage{colordvi} 
\usepackage{bm}
\usepackage{epsfig}
\usepackage{hyperref}
\usepackage[dvipsnames]{xcolor}
\usepackage{color, colortbl}
\usepackage[T1]{fontenc} 
\usepackage{braket}
\usepackage{euclid}
\usepackage{placeins}
\usepackage{tabularx}
\usepackage{comment}
\graphicspath{{./Figures/}}
\usepackage[normalem]{ulem}

\definecolor{amaranth}{rgb}{0.9, 0.17, 0.31}
\definecolor{forestgreen(web)}{rgb}{0.13, 0.55, 0.13}
\definecolor{lavender(web)}{rgb}{0.9, 0.9, 0.98}
\definecolor{cosmiclatte}{rgb}{1.0, 0.97, 0.91}
\definecolor{jonquil}{rgb}{0.98, 0.85, 0.37}
\definecolor{khaki(x11)(lightkhaki)}{rgb}{0.94, 0.9, 0.55}
\definecolor{thistle}{rgb}{0.85, 0.75, 0.85}

\newcommand{\GCsp}{\text{GC}\ensuremath{_\mathrm{sp}}}
\newcommand{\GCph}{\text{GC}\ensuremath{_\mathrm{ph}}}
\newcommand{\Omegam}{\ensuremath{\Omega_{\mathrm{m},0}}}

\newcommand{\lcdm}{\ensuremath{\Lambda\mathrm{CDM}}}

\newcommand{\de}{\mathrm{d}}

\defcitealias{Euclid:2019clj}{EC19}

\usepackage[nameinlink,noabbrev]{cleveref}
\Crefname{equation}{Eq.}{Eqs.}
\Crefname{section}{Sect.}{Sects.}
\Crefname{figure}{Fig.}{Figs.}
\Crefname{equation}{Equation}{Equations}
\Crefname{section}{Section}{Sections}
\Crefname{figure}{Figure}{Figures}

\begin{document}
\title{\Euclid: 
Constraining linearly scale-independent modifications of gravity with the spectroscopic and photometric primary probes\thanks{This paper is published on behalf of the Euclid Consortium.}}


\newcommand{\orcid}[1]{} 
\author{N.~Frusciante$^{1}$\thanks{\email{noemi.frusciante@unina.it}}, F.~Pace\orcid{0000-0001-8039-0480}$^{2,3,4}$, V.F.~Cardone$^{5,6}$, S.~Casas\orcid{0000-0002-4751-5138}$^{7}$, I.~Tutusaus\orcid{0000-0002-3199-0399}$^{8,9,10,11}$, M.~Ballardini\orcid{0000-0003-4481-3559}$^{12,13,14,15}$, E.~Bellini $^{16,17,18,8}$, G.~Benevento\orcid{0000-0002-6999-2429}$^{19}$, B.~Bose\orcid{0000-0003-1965-8614}$^{20,8,21}$, P.~Valageas$^{22}$
% ...
}

\institute{$^{1}$ Department of Physics "E. Pancini", University Federico II, Via Cinthia 6, 80126, Napoli, Italy\\
$^{2}$ Dipartimento di Fisica, Universit\`a degli Studi di Torino, Via P. Giuria 1, 10125 Torino, Italy\\
$^{3}$ INFN-Sezione di Torino, Via P. Giuria 1, 10125 Torino, Italy\\
$^{4}$ INAF-Osservatorio Astrofisico di Torino, Via Osservatorio 20, 10025 Pino Torinese (TO), Italy\\
$^{5}$ INAF-Osservatorio Astronomico di Roma, Via Frascati 33, 00078 Monteporzio Catone, Italy\\
$^{6}$ INFN-Sezione di Roma, Piazzale Aldo Moro, 2 - c/o Dipartimento di Fisica, Edificio G. Marconi, 00185 Roma, Italy\\
$^{7}$ Institute for Theoretical Particle Physics and Cosmology (TTK), RWTH Aachen University, 52056 Aachen, Germany\\
$^{8}$ Universit\'e de Gen\`eve, D\'epartement de Physique Th\'eorique and Centre for Astroparticle Physics, 24 quai Ernest-Ansermet, CH-1211 Gen\`eve 4, Switzerland\\
$^{9}$ Institute of Space Sciences (ICE, CSIC), Campus UAB, Carrer de Can Magrans, s/n, 08193 Barcelona, Spain\\
$^{10}$ Institut d'Estudis Espacials de Catalunya (IEEC), Carrer Gran Capit\'a 2-4, 08034 Barcelona, Spain\\
$^{11}$ Institut de Recherche en Astrophysique et Plan\'etologie (IRAP), Universit\'e de Toulouse, CNRS, UPS, CNES, 14 Av. Edouard Belin, 31400 Toulouse, France\\
$^{12}$ Dipartimento di Fisica e Scienze della Terra, Universit\'a degli Studi di Ferrara, Via Giuseppe Saragat 1, 44122 Ferrara, Italy\\
$^{13}$ Dipartimento di Fisica e Astronomia, Universit\'a di Bologna, Via Gobetti 93/2, 40129 Bologna, Italy\\
$^{14}$ INAF-Osservatorio di Astrofisica e Scienza dello Spazio di Bologna, Via Piero Gobetti 93/3, 40129 Bologna, Italy\\
$^{15}$ INFN-Sezione di Bologna, Viale Berti Pichat 6/2, 40127 Bologna, Italy\\
$^{16}$ INFN, Sezione di Trieste, Via Valerio 2, 34127 Trieste TS, Italy\\
$^{17}$ IFPU, Institute for Fundamental Physics of the Universe, via Beirut 2, 34151 Trieste, Italy\\
$^{18}$ SISSA, International School for Advanced Studies, Via Bonomea 265, 34136 Trieste TS, Italy\\
$^{19}$ Johns Hopkins University 3400 North Charles Street Baltimore, MD 21218, USA\\
$^{20}$ Institute for Astronomy, University of Edinburgh, Royal Observatory, Blackford Hill, Edinburgh EH9 3HJ, UK\\
$^{21}$ Institute for Computational Science, University of Zurich, Winterthurerstrasse 190, 8057 Zurich, Switzerland\\
$^{22}$ Institut de Physique Th\'eorique, CEA, CNRS, Universit\'e Paris-Saclay 91191 Gif-sur-Yvette Cedex, France\\
$^{23}$ Dipartimento di Fisica e Astronomia "G.Galilei", Universit\`a di Padova, Via Marzolo 8, 35131 Padova, Italy\\
$^{24}$ INFN-Padova, Via Marzolo 8, 35131 Padova, Italy\\
$^{25}$ INAF-Osservatorio Astronomico di Padova, Via dell'Osservatorio 5, 35122 Padova, Italy\\
$^{26}$ CERN, Theoretical Physics Department, Geneva, Switzerland\\
$^{27}$ Department of Physics, Oxford University, Keble Road, Oxford OX1 3RH, UK\\
$^{28}$ INFN-Bologna, Via Irnerio 46, 40126 Bologna, Italy\\
$^{29}$ Institute of Cosmology and Gravitation, University of Portsmouth, Portsmouth PO1 3FX, UK\\
$^{30}$ Dipartimento di Scienze Matematiche, Fisiche e Informatiche, Universit\`a di Parma, Viale delle Scienze 7/A 43124 Parma, Italy\\
$^{31}$ INFN Gruppo Collegato di Parma, Viale delle Scienze 7/A 43124 Parma, Italy\\
$^{32}$ Instituto de Astrof\'isica e Ci\^encias do Espa\c{c}o, Faculdade de Ci\^encias, Universidade de Lisboa, Campo Grande, 1749-016 Lisboa, Portugal\\
$^{33}$ Institut d'Astrophysique de Paris, UMR 7095, CNRS, and Sorbonne Universit\'e, 98 bis boulevard Arago, 75014 Paris, France\\
$^{34}$ Institut f\"ur Theoretische Physik, University of Heidelberg, Philosophenweg 16, 69120 Heidelberg, Germany\\
$^{35}$ Universit\'e St Joseph; Faculty of Sciences, Beirut, Lebanon\\
$^{36}$ Institute Lorentz, Leiden University, PO Box 9506, Leiden 2300 RA, The Netherlands\\
$^{37}$ Institute of Theoretical Astrophysics, University of Oslo, P.O. Box 1029 Blindern, 0315 Oslo, Norway\\
$^{38}$ Universit\'e Paris-Saclay, CNRS, Institut d'astrophysique spatiale, 91405, Orsay, France\\
$^{39}$ Mullard Space Science Laboratory, University College London, Holmbury St Mary, Dorking, Surrey RH5 6NT, UK\\
$^{40}$ Dipartimento di Fisica, Universit\`a degli studi di Genova, and INFN-Sezione di Genova, via Dodecaneso 33, 16146, Genova, Italy\\
$^{41}$ INFN-Sezione di Roma Tre, Via della Vasca Navale 84, 00146, Roma, Italy\\
$^{42}$ INAF-Osservatorio Astronomico di Capodimonte, Via Moiariello 16, 80131 Napoli, Italy\\
$^{43}$ Instituto de Astrof\'isica e Ci\^encias do Espa\c{c}o, Universidade do Porto, CAUP, Rua das Estrelas, PT4150-762 Porto, Portugal\\
$^{44}$ INAF-IASF Milano, Via Alfonso Corti 12, 20133 Milano, Italy\\
$^{45}$ Institut de F\'{i}sica d'Altes Energies (IFAE), The Barcelona Institute of Science and Technology, Campus UAB, 08193 Bellaterra (Barcelona), Spain\\
$^{46}$ Port d'Informaci\'{o} Cient\'{i}fica, Campus UAB, C. Albareda s/n, 08193 Bellaterra (Barcelona), Spain\\
$^{47}$ INFN section of Naples, Via Cinthia 6, 80126, Napoli, Italy\\
$^{48}$ Dipartimento di Fisica e Astronomia "Augusto Righi" - Alma Mater Studiorum Universit\`a di Bologna, Viale Berti Pichat 6/2, 40127 Bologna, Italy\\
$^{49}$ INAF-Osservatorio Astrofisico di Arcetri, Largo E. Fermi 5, 50125, Firenze, Italy\\
$^{50}$ Centre National d'Etudes Spatiales, Toulouse, France\\
$^{51}$ Institut national de physique nucl\'eaire et de physique des particules, 3 rue Michel-Ange, 75794 Paris C\'edex 16, France\\
$^{52}$ European Space Agency/ESRIN, Largo Galileo Galilei 1, 00044 Frascati, Roma, Italy\\
$^{53}$ ESAC/ESA, Camino Bajo del Castillo, s/n., Urb. Villafranca del Castillo, 28692 Villanueva de la Ca\~nada, Madrid, Spain\\
$^{54}$ Univ Lyon, Univ Claude Bernard Lyon 1, CNRS/IN2P3, IP2I Lyon, UMR 5822, 69622, Villeurbanne, France\\
$^{55}$ Institute of Physics, Laboratory of Astrophysics, Ecole Polytechnique F\'{e}d\'{e}rale de Lausanne (EPFL), Observatoire de Sauverny, 1290 Versoix, Switzerland\\
$^{56}$ Departamento de F\'isica, Faculdade de Ci\^encias, Universidade de Lisboa, Edif\'icio C8, Campo Grande, PT1749-016 Lisboa, Portugal\\
$^{57}$ Department of Astronomy, University of Geneva, ch. d'Ecogia 16, 1290 Versoix, Switzerland\\
$^{58}$ Universit\'e Paris-Saclay, Universit\'e Paris Cit\'e, CEA, CNRS, Astrophysique, Instrumentation et Mod\'elisation Paris-Saclay, 91191 Gif-sur-Yvette, France\\
$^{59}$ INAF-Osservatorio Astronomico di Trieste, Via G. B. Tiepolo 11, 34143 Trieste, Italy\\
$^{60}$ Istituto Nazionale di Fisica Nucleare, Sezione di Bologna, Via Irnerio 46, 40126 Bologna, Italy\\
$^{61}$ Max Planck Institute for Extraterrestrial Physics, Giessenbachstr. 1, 85748 Garching, Germany\\
$^{62}$ Universit\"ats-Sternwarte M\"unchen, Fakult\"at f\"ur Physik, Ludwig-Maximilians-Universit\"at M\"unchen, Scheinerstrasse 1, 81679 M\"unchen, Germany\\
$^{63}$ Dipartimento di Fisica "Aldo Pontremoli", Universit\`a degli Studi di Milano, Via Celoria 16, 20133 Milano, Italy\\
$^{64}$ INFN-Sezione di Milano, Via Celoria 16, 20133 Milano, Italy\\
$^{65}$ INAF-Osservatorio Astronomico di Brera, Via Brera 28, 20122 Milano, Italy\\
$^{66}$ Jet Propulsion Laboratory, California Institute of Technology, 4800 Oak Grove Drive, Pasadena, CA, 91109, USA\\
$^{67}$ von Hoerner \& Sulger GmbH, Schlo{\ss}Platz 8, 68723 Schwetzingen, Germany\\
$^{68}$ Technical University of Denmark, Elektrovej 327, 2800 Kgs. Lyngby, Denmark\\
$^{69}$ Max-Planck-Institut f\"ur Astronomie, K\"onigstuhl 17, 69117 Heidelberg, Germany\\
$^{70}$ Aix-Marseille Universit\'e, CNRS/IN2P3, CPPM, Marseille, France\\
$^{71}$ Department of Physics and Helsinki Institute of Physics, Gustaf H\"allstr\"omin katu 2, 00014 University of Helsinki, Finland\\
$^{72}$ NOVA optical infrared instrumentation group at ASTRON, Oude Hoogeveensedijk 4, 7991PD, Dwingeloo, The Netherlands\\
$^{73}$ Argelander-Institut f\"ur Astronomie, Universit\"at Bonn, Auf dem H\"ugel 71, 53121 Bonn, Germany\\
$^{74}$ Dipartimento di Fisica e Astronomia "Augusto Righi" - Alma Mater Studiorum Universit\`{a} di Bologna, via Piero Gobetti 93/2, 40129 Bologna, Italy\\
$^{75}$ Department of Physics, Institute for Computational Cosmology, Durham University, South Road, DH1 3LE, UK\\
$^{76}$ European Space Agency/ESTEC, Keplerlaan 1, 2201 AZ Noordwijk, The Netherlands\\
$^{77}$ Department of Physics and Astronomy, University of Aarhus, Ny Munkegade 120, DK-8000 Aarhus C, Denmark\\
$^{78}$ Centre for Astrophysics, University of Waterloo, Waterloo, Ontario N2L 3G1, Canada\\
$^{79}$ Department of Physics and Astronomy, University of Waterloo, Waterloo, Ontario N2L 3G1, Canada\\
$^{80}$ Perimeter Institute for Theoretical Physics, Waterloo, Ontario N2L 2Y5, Canada\\
$^{81}$ Space Science Data Center, Italian Space Agency, via del Politecnico snc, 00133 Roma, Italy\\
$^{82}$ Institute of Space Science, Bucharest, 077125, Romania\\
$^{83}$ Instituto de Astrof\'isica de Canarias, Calle V\'ia L\'actea s/n, 38204, San Crist\'obal de La Laguna, Tenerife, Spain\\
$^{84}$ Departamento de Astrof\'{i}sica, Universidad de La Laguna, 38206, La Laguna, Tenerife, Spain\\
$^{85}$ Departamento de F\'isica, FCFM, Universidad de Chile, Blanco Encalada 2008, Santiago, Chile\\
$^{86}$ Aix-Marseille Universit\'e, CNRS, CNES, LAM, Marseille, France\\
$^{87}$ Centro de Investigaciones Energ\'eticas, Medioambientales y Tecnol\'ogicas (CIEMAT), Avenida Complutense 40, 28040 Madrid, Spain\\
$^{88}$ Instituto de Astrof\'isica e Ci\^encias do Espa\c{c}o, Faculdade de Ci\^encias, Universidade de Lisboa, Tapada da Ajuda, 1349-018 Lisboa, Portugal\\
$^{89}$ Universidad Polit\'ecnica de Cartagena, Departamento de Electr\'onica y Tecnolog\'ia de Computadoras, 30202 Cartagena, Spain\\
$^{90}$ Kapteyn Astronomical Institute, University of Groningen, PO Box 800, 9700 AV Groningen, The Netherlands\\
$^{91}$ Infrared Processing and Analysis Center, California Institute of Technology, Pasadena, CA 91125, USA\\
$^{92}$ Institut d'Astrophysique de Paris, 98bis Boulevard Arago, 75014, Paris, France}

\date{\today}

\authorrunning{N. Frusciante et al.}

\titlerunning{\Euclid forecasts on modified gravity}


\abstract
{The future \Euclid space satellite mission will offer an invaluable opportunity to constrain modifications to Einstein's general relativity at cosmic scales. 
In this paper, we focus on modified gravity models characterised, at linear scales, by a scale-independent growth of perturbations while featuring different testable types of derivative screening mechanisms at smaller non-linear scales.}
{We considered three specific models, namely Jordan-Brans-Dicke (JBD), a scalar-tensor theory with a flat potential, the normal branch of Dvali-Gabadadze-Porrati (nDGP) gravity, a braneworld model in which our Universe is a four-dimensional brane embedded in a five-dimensional Minkowski space-time, and $k$-mouflage (KM) gravity, an extension of $k$-essence scenarios with a universal coupling of the scalar field to matter. In preparation for real data, we provide forecasts from spectroscopic and photometric primary probes by \Euclid on the cosmological parameters and the additional parameters of the models, respectively, $\omega_{\rm BD}$, $\Omega_{\rm rc}$ and $\epsilon_{2,0}$, which quantify the deviations from general relativity. This analysis will improve our knowledge of the cosmology of these modified gravity models.}
{The forecast analysis employs the Fisher matrix method applied to weak lensing (WL); photometric galaxy clustering (\GCph), spectroscopic galaxy clustering (\GCsp) and the cross-correlation (XC) between \GCph\ and WL. 
For the \Euclid survey specifications, we define three scenarios that are characterised by different cuts in the maximum multipole and wave number, to assess the constraining power of non-linear scales. For each model we considered two fiducial values for the corresponding model parameter.} 
{In an optimistic setting at 68.3\% confidence interval, we find the following percentage relative errors with \Euclid alone: for $\log_{10}{\omega_{\rm BD}}$, with a fiducial value of $\omega_{\rm BD}=800$, 27.1\% using \GCsp\ alone, 3.6\% using \GCph+WL+XC and 3.2\% using \GCph+WL+XC+\GCsp; for $\log_{10}{\Omega_{\rm rc}}$, with a fiducial value of $\Omega_{\rm rc}=0.25$, we find  93.4\%, 20\% and 15\% respectively; and finally, for $\epsilon_{2,0}=-0.04$, we find 3.4\%, 0.15\%, and 0.14\%. 
From the relative errors for fiducial values closer to their \lcdm\ limits, we find that most of the constraining power is lost. Our results highlight the importance of the constraining power  from non-linear scales.}
{}


\maketitle

\section{Introduction} \label{sec:intro}
We have entered a new era in gravitational physics in which it is now possible to test and exploit general relativity (GR) on a wide range of scales. The successes of Solar System constraints, and the precision measurements arising from observations of millisecond pulsars can now be combined with detections of gravitational waves and images of black hole shadows \citep{Berti:2015itd}. To this battery of techniques  must be added cosmological constraints using the large-scale structure of the Universe \citep{Ferreira:2019xrr}.

There have been multiple attempts at constraining GR with cosmological surveys \citep[see, for example][]{Planck:2018vyg,Mueller:2016kpu,Joudaki:2017zdt,DES:2021zdr,Raveri:2021dbu,Nguyen:2023fip}. With spectroscopic and imaging surveys of galaxies, combined with measurements of the cosmic microwave background (CMB), it has been possible to map out gravitational potentials over an appreciable part of the Universe as well as determine the growth rate of the structure and its morphology out to redshift $z\sim 2$. The resulting constraints have been an important first step in understanding gravity in an altogether untested regime, but they have been underwhelming. For example, the constraints on the simplest scalar-tensor modification to GR, Jordan-Brans-Dicke (JBD) gravity, are more than order of magnitude weaker \citep{Joudaki:2020shz,Ballardini:2021evv} than those obtained from millisecond pulsar observations \citep{Voisin:2020lqi}, but on very different scales.

The relevance of cosmological measurements for gravitational physics is about to change with the upcoming generation of large-scale structure surveys \citep{Ferreira:2019xrr}. By mapping out vast swathes of the Universe with exquisite precision, it is hoped that it will be possible to substantially  tighten  the constraints on gravitational physics on the largest observable scales. Of particular importance in this new vanguard is the \Euclid mission. The \Euclid satellite will undertake two key complementary surveys: a spectroscopic survey of galaxies and an imaging survey (targeting weak lensing; it can also be used to reconstruct galaxy clustering using photometric redshifts). The primary goal is to determine the nature of dark energy and it is ideally suited for cosmological constraints on gravity \cite[][EC19 hereafter]{Euclid:2019clj}.

Given the potential of the \Euclid mission it is imperative to assess its ability to constrain GR. One way of doing this is by assessing how well it will be able to constrain specific extensions of GR, in particular, modified gravity models (MG). 
MG models are particularly constrained by small-scale experiments such as those in the Solar System where fifth-force effects \citep{Bertotti:2003rm} and a violation of the equivalence principle \citep{Williams:2012nc,Touboul:2017grn} have been thoroughly investigated. As a result, extensions of GR that preserve the equivalence principle in the Earth's environment and exclude large fifth-force effects on Solar System scales must shield small-scales from potential deviations from GR on large cosmological distances. New physical effects would only reveal themselves on large scales where the \Euclid mission would have the potential to unravel them. Screenings of GR extensions have been broadly classified into three families: the chameleon \citep{Khoury:2003aq}, $k$-mouflage \citep{Babichev_2009} and Vainshtein \citep{Vainshtein:1972sx} mechanisms \citep[see, for a recent review,][]{Brax:2021wcv}. For chameleons, screening occurs in regions of space where Newton's potential is large enough whilst for $k$-mouflage and Vainshtein, this takes place where the first or second spatial derivatives of Newton's potential are also large enough.
One particular example of an extensively studied theory with the chameleon screening property is $f(R)$ gravity \citep{Carroll:2003wy,Hu:2007nk}. It has been shown that the \Euclid mission, using the combination of spectroscopic and photometric probes, will be able to distinguish this model from \lcdm\ at more than $3\sigma$ confidence level, for realistic fiducial values of its free model parameter, $f_{R0}$ \citep{Casas:2023lnp}. 
Derivative screening mechanisms, that is, $k$-mouflage and Vainshtein, have not been investigated within the context of the \Euclid mission, and this will be one of the outcomes of the present work.

In this paper, we forecast how well the surveys from the \Euclid mission can be used to constrain a family of theories that modify the theory of GR, but retain one of its properties: a scale-independent linear growth rate. The three theories we  consider are: JBD gravity \citep{Brans:1961sx}, which is the simplest scalar-tensor theory and involves a non-minimal coupling between a scalar field and the metric; Dvali-Gabadadze-Porrati gravity \citep[DGP, see][]{Dvali:2000hr}, which is  a braneworld model that introduces modifications on cosmological scales and screens with the Vainshtein mechanism; and a $k$-mouflage model \citep[KM, see][]{Babichev_2009,Brax:2014yla,Brax_2016}, that is a scalar-tensor theory with a non-canonical kinetic energy and the $k$-mouflage screening property. 
The JBD theory is not screened, and  to be consistent with current astrophysical constraints on fifth forces,  it requires that the Brans-Dicke (BD) parameter $\omega_{\rm BD}>4\times 10^4$ \citep{Bertotti:2003rm}. This is larger than the range of parameters that can cosmologically be tested,  as we show below. It implies that the JBD model investigated here must be taken as a template for large-scale deviations against which we compare the DGP and $k$-mouflage models. All these modifications of GR affect in one way or another the expansion rate and growth of structure and they are therefore prime candidates to be constrained by data from the \Euclid mission.

We structure this paper as follows. In Sect.~\ref{sec:models} we recapitulate the essential facts about linear cosmological perturbations in the context of extensions to GR, identifying the phenomenological time-dependent parameters, ($\mu$, $\eta$, and $\Sigma$), that are fed into the linear evolution equations. We then describe the three candidate theories we explored by laying out their corresponding actions, background evolution, functional forms of \{$\mu$, $\eta$, and $\Sigma$\}, and how all this is implemented numerically. In Sect.~\ref{sec:thpred} we explain in detail  how we calculated all aspects of the theoretical model predictions that  go into the forecasting procedure. In Sect.~\ref{sec:fisher} we describe the survey specifications and how they are integrated in the analysis method; in this paper, we use a Fisher forecasting approach. In Sect.~\ref{sec:results} we present the results of our methods, and we conclude  in Sect.~\ref{sec:conclusions}.

\section{Linearly scale-independent modified gravity} \label{sec:models}

We followed the Bardeen formalism \citep{Bardeen:1980kt,1995ApJ...455....7M} and defined the infinitesimal line element of the flat Friedmann-Lemaître-Robertson-Walker (FLRW) metric by
\begin{equation}
\label{eq:perturbed_metric}
 \de s^2 = -(1+2\Psi)\,c^2\,\de t^2 + a^2(t)\,(1-2\Phi)\,\delta_{ij} \, \de x^i\,\de x^j\,,
\end{equation}
where $a(t)$ is the scale factor as a function of the cosmic time $t$, $\Psi$ and $\Phi$ are the two scalar potentials and $c$ is the speed of light. We  work in Fourier space, where modes are functions of $t$ and the comoving wave number $k^i$. At linear order, the matter energy-momentum tensor can be decomposed as 
\begin{equation}
T^0_0=-\bar{\rho}(1+\delta)c^2\,\,,\,
T^0_i=(\bar{\rho}+\bar{p}/c^2)cv_i\,\,,\,
T^i_j=(\bar{p}+\delta p)\delta^i_j+\Sigma^i_j\,, \label{eq:en_mom_tensor}
\end{equation}
where $\delta=\rho/\bar{\rho}-1$ is the energy density contrast with $\rho$ being the matter density and $\bar{\rho}$ its background value, $p=\bar{p}+\delta p$ is the pressure with $\bar{p}$ the background value, $v_i$ is the peculiar velocity and $\Sigma^i_j$ is the (traceless) anisotropic stress, $\Sigma^i_i=0$. In the following we work with the scalar component of the matter anisotropic stress and the comoving density perturbation,  defined as $(\bar{\rho}c^2+\bar{p})\left(\hat{k}^i\cdot\hat{k}_j-1/3\delta^i_j\right)\sigma=\Sigma^i_j$ with $\hat{k}_i=k_i/k$ and  $\bar\rho\Delta\equiv\bar\rho\delta+3(aH/k/c^2)(\bar{\rho}+\bar{p}/c^2)v$, where $v$ is the velocity potential defined through $v_i=-\imath k_i v/k$. Here $H \equiv \dot{a}/a$ is the Hubble function and a dot stands for the derivative with respect to the coordinate time. In analogy to the \lcdm\ model where curvature is found to be compatible with zero,  we  assumed a flat spatial geometry.

In models in which gravity is modified by  a scalar field, the relations between the gravitational potentials and the matter perturbations are modified. These deviations from GR can be encoded into two functions, defined as follows:
\begin{align}
-k^2\Psi&=\frac{4\pi\,G_{\rm N}}{c^2} \,a^2\mu(a,k)\left[\bar\rho\Delta+3\left(\bar\rho+\bar {p}/c^2\right)\sigma\right]\,, \label{eq:mu}\\ 
-k^2\left(\Phi+\Psi\right) & = \frac{8\pi\,G_{\rm N}}{c^2}\,a^2\Big\{\Sigma(a,k)\left[\bar\rho\Delta+3\left(\bar\rho+\bar{p}/c^2\right)\sigma\right] \nonumber \\
& -\frac{3}{2}\mu(a,k)\left(\bar\rho+\bar{p}/c^2\right)\sigma\Big\}\,, \label{eq:sigma}
\end{align}
where $G_{\rm N}$ is Newton's gravitational constant. Eqs.~(\ref{eq:mu}) and (\ref{eq:sigma}) can be obtained using the quasi-static approximation, which considers scales smaller  than the horizon and the sound-horizon of the scalar field, where time derivatives become negligible with respect to spatial derivatives. In a given model, it allows us to determine  the functional form of $\mu$ and $\Sigma$ analytically \citep[see, e.g.][]{Silvestri:2013ne,Bellini:2014fua,Gleyzes:2014rba,Zucca:2019xhg,Pace:2020qpj}. A further function that can be introduced is therefore the function that defines the ratio of the two potentials, $\eta=\Phi/\Psi$. In the absence of anisotropic stress the three phenomenological functions are related by the following expression:
\begin{equation}
 \Sigma = \frac{1}{2}\mu(1+\eta)\,.
\end{equation}

The \emph{phenomenological functions} $\mu$, $\eta$ and $\Sigma$ are identically equal to $1$ in the GR limit. In general, they are functions of time and scale. The models that we consider in this paper preserve the scale-independent (linear) growth pattern, that is, they have $\mu=\mu(a)$, $\Sigma=\Sigma(a)$ and $\eta=\eta(a)$. 
 
For a given theory, $\mu$ and $\Sigma$ can be determined numerically after solving for the full dynamics of linear perturbations via an Einstein-Boltzmann solver. This can be achieved with \texttt{hi\_class} \citep{Zumalacarregui:2016pph,Bellini:2019syt} or \texttt{EFTCAMB} \citep{Hu:2013twa,Raveri:2014cka}, which implement the effective field theory formalism for dark energy into the standard \texttt{CLASS} \citep{Blas:2011rf,Lesgourgues:2011re} and \texttt{CAMB} \citep{Lewis:1999bs} codes, respectively. These codes have been validated as part of an extended code comparison effort \citep{Bellini:2017avd}. 
Alternatively, one can opt for the quasi-static (QS) limit, that is, scales sufficiently small to be well within the horizon and the sound-horizon of the scalar field, and derive the phenomenological functions analytically. In this case, one can use the \texttt{MGCAMB} patch to \texttt{CAMB} \citep{Zhao:2008bn,Hojjati:2011ix,Zucca:2019xhg}.

In the following, we  introduce the three models under consideration, that is, JBD, DGP, and KM. We  provide the background evolution equations and the expressions for $\mu$, $\eta$, and $\Sigma$ for each of these models.

\subsection{Jordan-Brans-Dicke gravity}\label{Sec:theoryJBD}

The JBD theory of gravity \citep{Brans:1961sx} is described by the following action,
\begin{equation}
 S_{\rm BD} = \int \de^4 x\sqrt{-g}\left[\frac{c^4}{16\pi}\left({\phi}R
-\frac{\omega_{\rm BD}}{\phi}g^{\mu\nu}\partial_\mu\phi\partial_\nu\phi - 2\Lambda\right)+\mathcal{L}_{\rm m}\right]\,, 
\label{JBD1}
\end{equation}
where $g_{\mu\nu}$ is the metric and $R$ its associated Ricci scalar, $\phi$ is the JBD scalar field (which has the dimensions of the inverse of Newton's constant), $\omega_{\rm BD}$ is the dimensionless BD parameter, and $\mathcal{L}_{\rm m}$ is the matter Lagrangian which is minimally coupled to the metric. 
We have included a cosmological constant, $\Lambda$.
The model has only one free parameter, $\omega_{\rm BD}$; in the limit in which $\omega_{\rm BD}\rightarrow\infty$, the scalar field is frozen and we recover Einstein gravity.

The JBD theory of gravity is remarkably simple in that it depends on so few parameters. However, cosmological constraints on JBD gravity can have wider implications when we take the view that it is the long-wavelength, low-energy limit of more general scalar-tensor theories \citep[see][for example]{Joudaki:2020shz}. Furthermore, more general scalar theories may be endowed with gravitational screening, which appears on smaller scales, or alternatively, regions of high density, for example. As a consequence, local constraints will to some extent decouple from more global, or large scale, constraints. Thus, constraints on cosmological scales on JBD theory may cover a broad class of scalar-tensor theories, and furthermore, be independent from model-specific constraints on small-scales. Thus, while simple, JBD gravity is a powerful tool for constraining general classes of scalar-tensor theories using cosmological data.

The modified Einstein field equations are \citep{Clifton:2011jh}
\begin{align}
  G_{\mu \nu} = {\frac{8\pi}{c^4}} \frac{1}{\phi} T_{\mu \nu} + &\, \frac{\omega_{\rm{BD}}}{\phi^2}\left(\nabla_\mu\phi\nabla_\nu\phi - \frac{1}{2}g_{\mu\nu}\nabla_\alpha\phi\nabla^\alpha\phi\right)  \nonumber \\
  + &\, \frac{1}{\phi}\left[\nabla_\mu\nabla_\nu\phi-g_{\mu \nu} \left(\Box \phi+\Lambda\right)\right]\,.
 \label{einsteineqJBD}
\end{align}
Here, $T_{\mu \nu}$ is the total matter stress-energy tensor, while  $\Box$ denotes the d'Alembertian. On the background, these equations give 
\begin{align}\label{FRWJBD}
3 H^{2} = &\, {\frac{8\pi}{c^4}} \frac{\bar{\rho}}{\phi} - 3H \frac{\dot{\phi}}{\phi} + \frac{\omega_{\rm{BD}}}{2}\frac{\dot{\phi}^{2}}{\phi^2} +\frac{\Lambda}{\phi}\,, \\
2\dot{H} + 3H^{2} = &\, -{\frac{8\pi}{c^4}}\frac{\bar{p}}{\phi} - \frac{\omega_{\rm{BD}}}{2}\frac{\dot{\phi}^{2}}{\phi^2} - 2H \frac{\dot{\phi}}{\phi} - \frac{\ddot{\phi}}{\phi}+\frac{\Lambda}{\phi}\,,
\end{align}
and the scalar field equation of motion reads
\begin{equation}
  \Box \phi = {\frac{8\pi}{c^4}} \left({\frac{T}{3 + 2 \omega_{\rm{BD}}}}\right) - \frac{4\Lambda}{3+2\omega_{\rm{BD}}}\,, \label{sfeqJBD}
\end{equation}
where $T \equiv g^{\mu\nu}T_{\mu\nu}$ is the trace of the stress-energy tensor on the background.

The phenomenological QS functions in this theory read
\begin{align}
\Sigma = &\, \frac{1}{G_{\rm N}\phi}\,, \nonumber \\
 \mu = &\, \frac{4 + 2\omega_{\rm BD}}{3 + 2 \omega_{\rm BD} }\Sigma \,, \\
 \eta \equiv &\, \frac{\Phi}{\Psi} = \, \frac{1+\omega_{\rm BD}}{2+\omega_{\rm BD}} \,. \nonumber
\end{align}

Constraints on JBD gravity have been obtained by using a combination of different cosmological data sets and sampling over a different parameterisation of the BD parameter $\omega_{\rm BD}$.
For example, \citet{Avilez:2013dxa} imposed a flat prior on 
$-\log_{10}\left(\omega_{\rm BD}\right)$ to obtain a lower bound of $\omega_{\rm BD} > 1900$ at 95\% CL with CMB information from {\em Planck} 2013 data.
\citet{Ballardini:2016cvy} obtained
$\log_{10}\left(1 + 1/\omega_{\rm BD}\right) < 0.0030$ at 95\% CL combining {\em Planck} 2015 and BOSS DR10-11 data; this upper bound was subsequently updated in \citet{Ballardini:2020iws} with a combination of {\em Planck} 2018 and Baryon Oscillation Spectroscopic Survey (BOSS) DR12 data to $\log_{10}\left(1 + 1/\omega_{\rm BD}\right) < 0.0022$ at 95\% CL.
\citet{Joudaki:2020shz} used a combination of the {\em Planck} 2018 CMB data, the $3\times2$pt combination of the Kilo-Degree Survey (KiDS) and the 2 degree Field gravitational Lens survey (2dFLens) data, the Pantheon supernovae data  and BOSS measurements of the BAO, to find the coupling constant $\omega_{\rm BD} > \{1540,\, 160,\, 160,\, 350\}$ at 95\% CL for the different choices of parametrization (or priors): $\left\{\log_{10}\left(1/\omega_{\rm BD}\right),\, \log_{10}\left(1+1/\omega_{\rm BD}\right),\, 1/\omega_{\rm BD},\, 1/\log_{10}\omega_{\rm BD}\right\}$. 
These constraints were obtained by fixing the value of the scalar field today to 
\begin{equation}
 G_{\rm N}\phi\left(a=1\right) \,= \, \frac{2\omega_{\rm BD}+4}{2\omega_{\rm BD}+3}\,,
\end{equation}
in order to guarantee that the effective gravitational constant at present corresponds to the one measured in a Cavendish-like experiment \citep{Boisseau:2000pr} (see also \cite{Avilez:2013dxa,Joudaki:2020shz,Ballardini:2021evv} for studies of the JBD model without imposing this condition and \citet{Ballardini:2016cvy,Ooba:2017gyn,Rossi:2019lgt,Braglia:2020iik,Braglia:2020auw,Cheng:2021yvh} for a simple generalisation of these constraints extending the JBD action (\ref{JBD1}) with different potentials and couplings to the Ricci scalar).

The JBD model is implemented in \texttt{CLASSig} \citep{Umilta:2015cta} and \texttt{hi\_class} \citep{Zumalacarregui:2016pph,Bellini:2019syt}. For these codes, the agreement and validation of the background and linear perturbations was thoroughly studied in \citet{Bellini:2017avd}.
In this paper, we use the results produced with \texttt{hi\_class}.

\subsection{Dvali-Gabadadze-Porrati braneworld gravity}\label{Sec:theorynDGP}

The DGP model~\citep{Dvali:2000hr} is a five-dimensional braneworld model defined by the action 
\begin{equation}
S = \frac{c^4}{16\pi G_5}\int_{\cal M} \de^5 x \sqrt{-\gamma} R_5
+ \int_{\partial {\cal M}} \de^4 x \sqrt{-g} 
\left[\frac{c^4}{16\pi G_{\rm N}} R + {\cal L}_{\rm m} \right]\,,
\end{equation}
where $\gamma$ is the five-dimensional metric and $R_5$ its Ricci curvature scalar. $G_5$ and $G_{\rm N}$ are the five- and four-dimensional Newton constants, respectively. 
The matter Lagrangian is denoted with ${\cal L}_{\rm m}$ and is confined to a four-dimensional brane in a five-dimensional Minkowski spacetime. The induced gravity described by the usual four-dimensional Einstein-Hilbert action is responsible for the recovery of the four-dimensional gravity on the brane. The cross-over scale $r_{\rm c} = G_5/(2 G_{\rm N})$ is the only parameter of the model and GR is recovered in the limit $r_{\rm c} \to \infty$. 

The Friedmann equation on the brane is given by~\citep{Deffayet:2000uy}
\begin{equation}
 H^2 = \pm\, c \frac{H}{r_{\rm c}} + \frac{8 \pi G}{3} \bar{\rho}\,.
\end{equation}
Two branches of solutions depend on the embedding of the brane: the self-accelerating branch  \citep[sDGP,][]{Bowcock:2000cq,Deffayet:2000uy} and the normal branch \citep[nDGP,][]{Bowcock:2000cq,Deffayet:2000uy}, corresponding to the $+$ and $-$ sign for the contribution from the five-dimensional gravity, respectively. The self-accelerating branch admits the late-time acceleration without dark energy but the solution is plagued by ghost instabilities \citep{Luty:2003vm,Gorbunov:2005zk,Charmousis:2006pn}. We therefore focus on the normal branch. In order to separate the effect of MG on structure formation from that on the expansion history, it is often assumed that the background expansion is identical to that of \lcdm. This can be achieved by introducing an additional dark energy contribution with an appropriate equation of state \citep{Schmidt:2009sv,Bag:2018jle}. We  adopted this approach. 

The evolution of density and metric perturbations on the brane require solutions of the bulk metric equations.
These bulk effects can be encapsulated in an effective $3+1$ description that uses the combination of any two of the functions $\mu$, $\eta$, and $\Sigma$.
Using the results of \cite{Koyama:2005kd,Hu:2007pj,Song:2007wd,Cardoso:2007xc,Lombriser:2009xg,Seahra:2010fj}, we have
\begin{equation}\label{eqn:eta}
 \eta = \frac{1+g}{1-g} \,,
\end{equation}
where, using the quasi-static (QS) approximation, we have \citep{Lombriser:2013aj}
\begin{equation}
 g(a) = g_{\rm QS} = -\frac{1}{3}\left[ 1 \mp \frac{2 H r_c}{c} \left( 1 + \frac{\dot{H}}{3H^2} \right) \right]^{-1} \,,
\end{equation}
so that the effective modification introduced with $\eta$ can be treated as scale-independent.

The effective $3+1$ Poisson equation for the lensing potential in the QS approximation is
\begin{equation}
- k^2 (\Phi + \Psi) = \frac{8\pi G_{\rm N}}{c^2}\,a^2\,\bar{\rho}\, \Delta\,, 
\end{equation}
therefore \citep{Lombriser:2013aj} 
\begin{equation}\label{eqn:sigmaDGP}
 \Sigma = 1\,,
\end{equation}
and hence
\begin{equation}\label{eqn:mu}
 \mu(a) = 1 + \frac{1}{3\beta}  \,, \qquad 
 \beta(a) \equiv 1+\frac{H}{H_0} \frac{1}{\sqrt{\Omega_{\rm rc}}}   \left(1+\frac{\dot{H}}{3H^2}\right) \,.
\end{equation}

We chose to parametrise the modification to gravity by $\Omega_{\rm rc} \equiv c^2/(4r_{\rm c}^2H_0^2)$.

Not many studies have constrained nDGP with an exact \lcdm\ background. However, \cite{2013MNRAS.436...89R}, using measurements of the zeroth- and second-order moments of the correlation function from SDSS DR7 data up to $r_{\rm max} = 120\,{\rm Mpc}\,h^{-1}$, and marginalised bias, found an $\Omega_{\rm rc}$ upper limit at 95\% of $\sim 40$ (from $r_{\rm c}>340$\,Mpc with fixed $H_0=70\,{\rm km}\,\rm{s}^{-1}\,{\rm Mpc}^{-1}$). In addition, \cite{Barreira:2016ovx} used the clustering wedges statistic of the galaxy correlation function and the growth rate values estimated from more recent BOSS DR12 data to set $[r_{\rm c}H_0/c]^{-1} < 0.97$ at 95\% C.L., corresponding to an upper bound of $\Omega_{\rm rc}< 0.27$.

There have also been recent attempts to forecast  constraints on $\Omega_{\rm rc}$. \cite{Liu:2021weo}, using the galaxy cluster abundance from the Simons Observatory and galaxy correlation functions from a Dark Energy Spectroscopic Instrument (DESI)-like experiment and found $\delta(\Omega_{\rm rc}) \sim$ 0.038 around a fiducial value of 0.25. \cite{Cataneo:2021xlx} forecast for \Euclid-like future constraints on a fiducial $\Omega_{\rm rc} \sim 0.0625$ a $1\sigma$ accuracy of 0.0125 from combining the 3D matter power spectrum and the probability distribution function of the smoothed three-dimensional matter density field probes.  \cite{Bose:2020wch} also forecast for a Large Synoptic Survey Telescope (LSST)-like survey a $1\sigma$ accuracy of 0.08 using cosmic shear alone on a fiducial $\Omega_{\rm rc} \sim 0$.

Constraints have also been inferred for an nDGP model with a cosmological constant rather than a constructed dark energy field, thus with an approximate \lcdm\ background.
For this model, \cite{Lombriser:2009xg} and \cite{Xu:2013ega} inferred somewhat stronger constraints of $\Omega_{\rm rc} < 0.020$ (95\%~C.L.) and $\Omega_{\rm rc} < 0.002$ (68\%~C.L.) from the combination of CMB and large-scale structure data, where the constraints were mainly driven by the CMB data.
Thus, high-precision CMB measurements such as by the Planck satellite  constrain $\Omega_{\rm rc}$ tightly. 
In this work, we focus on the model with an exact \lcdm\ background.

nDGP has been implemented in \texttt{MGCAMB} and in \texttt{QSA\_CLASS} \citep{Pace2021}. These codes solve for a \lcdm-background evolution and Eqs.~(\ref{eqn:sigmaDGP}) and (\ref{eqn:mu}). The overall agreement in the linear matter power spectrum is never worse than $0.5\%$. In this paper, we use the results produced with \texttt{MGCAMB}.

\subsection{\texorpdfstring{$k$}{k}-mouflage gravity}\label{Sec:theoryKM}

KM theories are built complementing simple $k$-essence scenarios with a universal coupling of the scalar field $\varphi$ to matter. They are defined by the action \citep{Babichev_2009,Brax:2014yla,Brax_2016}
\begin{align}
S & = \int \de^4 x \sqrt{-\tilde{g}} \left[ \frac{c^4}{16\pi \tilde{G}_{\rm N}} \tilde{R} + c^2 {\cal M}^4 K(\tilde{\chi}) \right] + S_{\rm m}(\psi_i, g_{\mu\nu})\,,
\label{eq:actkm}
\end{align}
where $\tilde{G}_{\rm N}$ is the bare Newton constant,  $\tilde{R}$ is the Ricci scalar in the Einstein frame, ${\cal M}^4$ is the energy density scale of the scalar field, $g_{\mu\nu}$ is the Jordan frame metric, $\tilde{g}_{\mu \nu}$ is the Einstein frame metric with $g_{\mu \nu}=A^2(\varphi)\tilde{g}_{\mu \nu}$, $S_{\rm m}$ is the Lagrangian of the matter fields $\psi_{m}^{(i)}$, $\tilde{\chi}$ is defined as
\begin{equation}
\tilde{\chi} = - \frac{\tilde{g}^{\mu\nu} \partial_{\mu}\varphi\partial_{\nu}\varphi}{2 {\cal M}^4} \,,
\end{equation}
and ${\cal M}^4 K$ is the non-standard kinetic term of the scalar field.

In these theories, the evolution of both the background and perturbations is affected by the universal coupling and by the scalar field dynamics. The degree of deviation from \lcdm\ at the background level and in perturbation theory can be expressed in terms of two time-dependent functions, related to the coupling $A$ and the kinetic function $K$, that is,
\begin{equation}
\label{epsilon2-def}
\epsilon_2\equiv \frac{\de \ln \bar{A}}{\de \ln a} \,, \quad \epsilon_1\equiv \frac{2}{\bar{K}'} \left[\epsilon_2  \frac{c}{\sqrt{8\pi\tilde{G}_{\rm N}}}  \left(\frac{\de \bar{\varphi}}{\de \ln a}\right)^{-1} \right]^2 \,, 
\end{equation}
where over-bars denote background quantities and a prime indicates derivatives with respect to $\bar{\tilde\chi}$. 
The KM Friedmann equation therefore reads
\begin{equation}
H^2 = \frac{8\pi\tilde{G}_{\rm N}}{3} \frac{\bar{A}^2}{(1-\epsilon_2)^2} \;
\left[\bar{\rho} + \frac{{\cal M}^4}{\bar{A}^4} \left(2 \bar{\tilde\chi} \frac{\de \bar{K}}{\de \bar{\tilde{\chi}}} - \bar{K}\right)\right] \,.
\label{E00}
\end{equation}
 
Considering linear scalar perturbations around an FLRW background, the phenomenological functions $\mu$ and $\Sigma$ read \citep{Benevento_2019}
\begin{equation}
\label{mu_Sigma_KM}
\mu(a) = (1+ \epsilon_1) \bar{A}^2 \,, \quad 
\Sigma(a) = \bar{A}^2 \,. 
\end{equation}
We recover GR when $\epsilon_1\to 0$, $\epsilon_2\to 0$, and $A$ and $K$ are constants, with ${\cal M}^4 K$ playing the role of the cosmological constant, see Eq.~(\ref{eq:actkm}).

In addition to the six standard \lcdm\ parameters, the KM model requires the specification of the kinetic function $K(\tilde\chi)$ and of the coupling $A(\varphi)$.
Following \cite{Brax_2016}, this can also be expressed in terms of the time-dependent background values $\bar K(a)$ and $\bar A(a)$.
\cite{Brax_2016} proposed a simple parameterisation that satisfies self-consistency constraints.
It involves five parameters, $\{\epsilon_{2,0},\gamma_A, m, \alpha_U,\gamma_U\}$, where $\epsilon_{2,0}$ is the value of $\epsilon_2$ at redshift $z=0$ and must be negative to ensure that there are no ghosts in the theory. For the coupling function $\bar A(a)$, this reads 
\begin{equation}
\label{Aa_KM}
\bar A(a) = 1 + \alpha_A - \alpha_A \left[ \frac{(\gamma_A+1)\, a}{\gamma_A+a} \right]^{\nu_A} \,, 
\end{equation}
with
\begin{equation}
\nu_A = \frac{3 (m-1)}{2 m - 1} \,, \quad \alpha_A = - \frac{\epsilon_{2,0}\, (1+\gamma_A)}{\gamma_A \nu_A}  \,. 
\end{equation}
Here, $m$ is the exponent of the kinetic function for a large argument, $K \sim \tilde{\chi}^m$, which also specifies the high-redshift dependence of $\bar A(a)$, while $\gamma_A$ sets the redshift at which $\bar{A}$ changes from the current unit value to the high-redshift value $1+\alpha_A$, which is parameterised by $\epsilon_{2,0}$. 
The background kinetic function $\bar{K}(a)$ is conveniently parameterised in terms of a function $U(a)$, with
\begin{equation}
\label{KpUa_KM}
\frac{\de\bar K'}{\de\tilde\chi} = \frac{U(a)}{a^3 \sqrt{\tilde\chi}} \, , \quad \sqrt{\tilde\chi} = - \frac{\bar\rho_0}{{\cal M}^4} \frac{\epsilon_2 \bar A^4}{2 U \left( - 3 \epsilon_2 + \frac{\de\ln U}{\de\ln a} \right)} \,,
\end{equation}
which we took to be of the form
\begin{equation}
\label{Ua_KM}
U(a) \propto \frac{a^2 \ln(\gamma_U+a)}{(\sqrt{a_{\rm eq}}+\sqrt{a}) \ln(\gamma_U+a)+\alpha_U a^2} \,.
\end{equation}
The various terms in this expression allowed us to follow the radiation, matter, and dark-energy eras.
The parameters $\alpha_U$ and $\gamma_U$, of order unity, set the shape of the transition to the dark-energy era.
The main parameter is $\epsilon_{2,0}$, which measures the amplitude of the deviation from GR and \lcdm\ at $z=0$. Other parameters, of order unity, mostly describe the shapes of the transitions between different cosmological regimes.

The KM model has been tested against a set of complementary cosmological data sets, including CMB temperature, polarisation and lensing, type Ia supernovae, baryon acoustic oscillations (BAO) and local measurements of $H_0$ \citep{Benevento_2019}. This gives the $95\%$ C.L. bounds $-0.04 \leq \epsilon_{2,0} \leq 0$, while the other parameters are unconstrained (as long as they do not become very large).
These results are consistent with earlier more qualitative CMB studies \citep{Barreira:2014gwa}.
The X-ray cluster multiplicity function could provide bounds of the same order \citep{Brax:2015lra}, but more detailed analyses are needed to derive robust constraints.

The fact that $\epsilon_{2,0}$ is the main parameter constrained by the data can be understood from Eqs.~(\ref{epsilon2-def}) and~(\ref{mu_Sigma_KM}) and the property that at leading order over $\epsilon_2$, we have $\epsilon_1 \simeq - \epsilon_2$. This shows that the running of Newton's constant and the impact on the linear metric and matter density fluctuations are set by $\epsilon_2$. Because most of the deviations from \lcdm\ occur at low redshift, this is mostly determined by the value $\epsilon_{2,0}$ at $z=0$. 
Therefore,  we here focused on the dependence on $\epsilon_{2,0}$, which was left free to vary, while we fixed the other parameters to $\gamma_A=0.2$, $m=3$, $\alpha_U=0.2$, and $\gamma_U=1$, as some representative values for the other parameters. 
Thus, this gives one additional parameter, $\epsilon_{2,0}$, in addition to the six standard \lcdm\ parameters.
The \lcdm\ model is recovered when $\epsilon_{2,0}=0$, independently of the value of the other parameters;  for $\epsilon_{2,0} \rightarrow 0$, we have:
\begin{equation}
\bar{A}(a)\rightarrow 1\,, \quad \epsilon_{2}(a)\rightarrow 0\,, \quad \epsilon_{1}(a)\rightarrow 0\,, \quad \forall \ a\,, \label{eq:LCDMlimit}
\end{equation}
and the kinetic function in Eq.~(\ref{eq:actkm}) reduces to a cosmological constant. 

To produce linear predictions for this theory we used its implementation in the \texttt{EFTCAMB} code that is described in \cite{Benevento_2019}.

\section{Theoretical predictions for \Euclid observables} \label{sec:thpred}

Both the forecasting method and the tools used in this paper are the same as those in \citetalias{Euclid:2019clj} except for the changes needed to account for the use of MG rather than GR in the predictions of \Euclid observables. The motivations and the relevant changes in the recipe have been described in a previous paper on forecasts for $f(R)$ theories. We therefore refer  to \cite{Casas:2023lnp}, while here we only recall the main steps.

\subsection{Photometric survey} \label{sec:photo}

The observables we considered for the \Euclid photometric survey are the angular power spectra $C_{ij}^{XY}(\ell)$ between the probe $X$ in the $i$\,-\,th redshift bin, and the probe $Y$ in the $j$\,-\,th bin, where $X$ refers to weak lensing (WL), photometric galaxy clustering (\GCph), or their cross correlation, XC. These are still calculated as in \citetalias{Euclid:2019clj} relying on the Limber approximation and setting to unity the $\ell$\,-\,dependent factor in the flat-sky limit. The spectra are then given by
\begin{equation}
 C^{XY}_{ij}(\ell) = c\int_{z_{\rm min}}^{z_{\rm max}}\de z\,{\frac{W_i^X(z)W_j^Y(z)}{H(z)r^2(z)}P_{\delta\delta}(k_\ell,z)}\,, \label{eq:ISTrecipe}
\end{equation}
with $k_\ell=(\ell+1/2)/r(z)$, $r(z)$ the comoving distance to redshift $z=1/a-1$, and $P_{\delta\delta}(k_\ell,z)$ the non-linear power spectrum of matter density fluctuations, $\delta$, at wave number $k_{\ell}$ and redshift $z$. We set $(z_{\rm min}, z_{\rm max}) = (0.001, 4)$, which spans the full range within which the source redshift distributions $n_i(z)$ are non-vanishing. The \GCph\ and WL window functions read \citep{SpurioMancini:2019rxy}
\begin{align}
 W_i^{\rm G}(k,z) =&\; b_i(k,z)\,\frac{n_i(z)}{\bar{n}_i}\frac{H(z)}{c}\,, \label{eq:wg_mg}\\  
 W_i^{\rm L}(k,z) =&\; \frac{3}{2}\Omegam \frac{H_0^2}{c^2}(1+z)\,r(z)\,\Sigma(z)
\int_z^{z_{\rm max}}{\de z'\frac{n_i(z')}{\bar{n}_i}\frac{r(z'-z)}{r(z')}}\nonumber\\
  &+W^{\rm IA}_i(k,z)\,, \label{eq:wl_mg}
\end{align}
where $n_i(z)/\bar{n}_i$ and $b_i(k,z)$ are the normalised galaxy distribution and the galaxy bias in the $i$-th redshift bin, respectively, and $W^{\rm IA}_i(k,z)$ encodes the contribution of intrinsic alignments (IA) to the WL power spectrum. The  function $\Sigma(z)$ in the WL window function explicitly accounts for the changes in the lensing potential due to the particular MG theory of interest. Its explicit form for the cases under consideration can be found in Section \ref{sec:models}. The impact on the background quantities $H(z)$, $r(z)$ and the matter power spectrum $P_{\delta \delta}(k, z)$, in contrast, are already taken into account by the dedicated Boltzmann solver so that the \GCph\ window function remains unchanged.

The IA contribution was computed following the eNLA model adopted in  \citetalias{Euclid:2019clj} so that the corresponding window function is
\begin{equation}\label{eq:IA}
 W^{\rm IA}_i(k,z)=-\frac{\mathcal{A}_{\rm IA}\,\mathcal{C}_{\rm IA}\,\Omega_{\rm m,0}\,\mathcal{F}_{\rm IA}(z)}{\delta(k,z)/\delta(k,z=0)}\frac{n_i(z)}{\bar{n}_i(z)}\frac{H(z)}{c}\,,
\end{equation}
where 
\begin{equation}
 \mathcal{F}_{\rm IA}(z)=(1+z)^{\eta_{\rm IA}}\left[\frac{\langle L\rangle(z)}{L_\star(z)}\right]^{\beta_{\rm IA}}\,,
\end{equation}
with $\langle L\rangle(z)$ and $L_\star(z)$ redshift-dependent mean and the characteristic luminosity of source galaxies as computed from the luminosity function, $\mathcal{A}_{\rm IA}$, $\beta_{\rm IA}$ and $\eta_{\rm IA}$ are the nuisance parameters of the model, and $\mathcal{C}_{\rm IA}$ is a constant accounting for dimensional units. This model is the same as was used in  \citetalias{Euclid:2019clj} since IA takes place on astrophysical scales that are unaffected by modifications to gravity. However, MG has an impact on the growth factor, introducing a possible scale dependence. This is explicitly taken into account in Eq.~(\ref{eq:IA}) through the matter perturbation $\delta(k,z)$, which is considered to be scale dependent in this case. This allows us to  also consider the scale dependence introduced by massive neutrinos, which was assumed to be negligible in \citetalias{Euclid:2019clj}. We nevertheless stress that for the models we considered, the scale dependence is quite small so that the IA is essentially the same as in the GR case.

\subsection{Spectroscopic survey}\label{sec:spect}

We now discuss the modelling of the power spectrum to analyse the data from the \Euclid spectroscopic survey.

For the models considered in this paper, the Compton wavelength of the scalar field is much larger than the scales probed by \Euclid. Moreover, it is assumed that the speed of propagation of the scalar fluctuations is of the order of the speed of light, so that the sound horizon is of the order of the Hubble scale. Under these assumptions, we can apply the quasi-static approximation and relate the scalar field perturbation to the gravitational potential. Since in all these models the weak equivalence principle holds, the modelling of the bias as an expansion in the derivatives of the gravitational potential remains unchanged with respect to the \lcdm\ one. To be consistent with the official forecast, we  used the same modelling for galaxy clustering as in \citetalias{Euclid:2019clj}. 

The observed galaxy power spectrum is given by 
\begin{multline}
P_\text{obs}(k, \mu_{\theta} ;z) = 
\frac{1}{q_\perp^2(z)\, q_\parallel(z)} 
\left\{\frac{\left[b\sigma_8(z)+f\sigma_8(z)\mu_{\theta}^2\right]^2}{1+ \Big[ f(z) k \mu_{\theta} \sigma_{\rm p}(z) \Big]^2 } \right\} 
\\
\times \frac{P_\text{dw}(k,\mu_{\theta};z)}{\sigma_8^2(z)}  
F_z(k,\mu_{\theta};z) 
+ P_\text{s}(z) \,, 
\label{eq:GC:pk-ext}
\end{multline}
where $P_{\rm dw}(k,\mu_{\theta};\,z)$, the de-wiggled power spectrum, includes the correction that accounts for the smearing of the BAO features,
\begin{equation}
P_\text{dw}(k,\mu_{\theta};z) = P_{\delta\delta}^{\rm lin}(k;z)\,\text{e}^{-g_\mu k^2} + P_\text{nw}(k;z)\left(1-\text{e}^{-g_\mu k^2}\right) \,,
\label{eq:pk_dw}
\end{equation}
and $P_{\delta\delta}^{\rm lin}(k;z)$ stands for the linear matter power spectrum. $P_{\rm nw}(k;z)$ is a no-wiggle power spectrum with the same broad-band shape as $P_{\delta\delta}^{\rm lin}(k;z)$ but without BAO features (see the discussion below). The function 
\be
g_\mu(k,\mu_\theta,z) = \sigma_{\rm v}^2(z) \left\{ 1 - \mu_\theta^2 + \mu_\theta^2 [1 + f(z)]^2  \right\}\,,
\ee
is the non-linear damping factor of the BAO signal derived by \citet{Eisenstein:2006nj}, with
\be
\sigma_{\rm v}^2 (z) = \frac{1}{6\pi^2} \int \de k P_{\delta \delta }^{\rm lin} (k,z) \,. \label{eq:sigmav}
\ee
The curly bracket in Eq. (\ref{eq:GC:pk-ext}) is the redshift-space-distortion (RSD) contribution correcting for the non-linear finger-of-God (FoG) effect, where we defined $b\sigma_8(z)$ as the product of the effective scale-independent bias of galaxy samples and the r.m.s.\ matter density fluctuation $\sigma_8(z)$ (we marginalized over $b\sigma_8(z)$), while $\mu_{\theta}$ is the cosine of the angle $\theta$ between the wave vector $\bm k$ and the line-of-sight direction $\hat{\bm r}$ and $\sigma_{\rm p}^2(z) = \sigma_{\rm v}^2(z)$. Although these parameters were assumed to be the same, they come from two different physical effects, namely large-scale bulk flow for the former and virial motion for the latter. 

The factor $F_z$ accounts for the smearing of the galaxy density field along the line of sight due to redshift uncertainties. It is given by
\begin{equation}
    F_z(k, \mu_{\theta};z) = \text{e}^{-k^2\mu_{\theta}^2\sigma_{r}^2(z)}\,,
\end{equation}
where $\sigma_{r}(z) = c\,(1+z)\sigma_{0,z}/H(z)$ and $\sigma_{0,z}$ is the error on the measured redshifts.

The factor in front of the curly bracket in Eq.~ (\ref{eq:GC:pk-ext}) describes the Alcock-Paczynski effect, which is parametrised in terms of the angular diameter distance $D_{\rm A}(z)$ and the Hubble parameter $H(z)$ as 
\begin{align}
q_{\perp}(z) &= \frac{D_{\rm A}(z)}{D_{\rm A,\, ref}(z)},\\
q_{\parallel}(z) &= \frac{H_\text{ref}(z)}{H(z)}\,.
\end{align}
Due to the Alcock-Paczynski effect, $\mu_{\theta}$ and $k$ were rescaled in a cosmology-dependent way as a function of the projection along and perpendicular to the line of sight. In the previous Eq.~(\ref{eq:GC:pk-ext}), the arguments $\mu$ and $k$ are themselves functions of the true $\mu_{\theta, \rm ref}$ and $k_{\rm ref}$ at the reference cosmology. This relation, which for each argument, is a function of $q_\parallel$ and $q_\perp$, can be found in Sect.~3.2.1 of \citetalias{Euclid:2019clj}.
Finally, the $P_{\rm s}(z)$ is a scale-independent shot-noise term that enters as a nuisance parameter \citepalias[see][]{Euclid:2019clj}.

Due to the scale independence of $\sigma_{\rm v} (= \sigma_{\rm p})$, we evaluated it at each redshift bin but we kept it fixed in the Fisher matrix analysis. This method corresponds to the optimistic settings in \citetalias{Euclid:2019clj}.
We would like to highlight that  we directly took  the derivatives of the observed galaxy power spectrum with respect to the final parameters, in contrast to \citetalias{Euclid:2019clj} where first a Fisher matrix analysis was performed for the redshift-dependent parameters $H(z)$, $D_{\rm A}(z)$, and $f\sigma_8(z)$ and then projected to the final cosmological parameters of interest. However, we verified that both approaches lead to consistent results when considering the $\Lambda$CDM and $w_0w_a$CDM models.
The other term appearing in Eq.~(\ref{eq:pk_dw}) is the non-wiggle matter power spectrum, and this was obtained applying a
Savitzky-Golay filter to the matter power spectrum $P_{\delta\delta}^{\rm lin}(k;z)$. For more details on this implementation, we refer to \citetalias{Euclid:2019clj}.

\subsection{non-linear modelling for weak lensing}\label{sec:non-linear-WL}

The galaxy power spectrum on mildly non-linear scales can be modelled by a modified version of the implementation in \citetalias{Euclid:2019clj}. This is no longer the case when we move to the deeply non-linear regime in which no an analytical description of the matter power spectrum is available for the models of our interest.

The wide window functions (in particular, the WL function) entering the prediction of the photometric observables $C_{ij}^{XY}(\ell)$ require an accurate description of non-linear scales. In particular,  the impact of baryons needs to be explicitely accounted for and nuisance parameters must be included,  which control the baryonic feedback in order to avoid biasing the parameter estimation \citep[see e.g.][]{Schneider:2019snl,Schneider:2019xpf}. However, we currently do not have accurate \Euclid-like simulations including baryonic effects, especially for MG cosmologies. Therefore, we ignored these effects in our analysis, and left their inclusion for a future work. We note that our scale cuts, in particular for the pessimistic scenario ($\ell < 1500 $), partially mitigate the fact of neglecting baryonic effects and are in agreement with recent studies on the impact of baryonic effects on constraints for modified gravity \citep[see, for instance][]{SpurioMancini:2023mpt}.
In the following subsections, we comment on the individual prescriptions for each MG model, wthat is used to compute the non-linear matter power spectrum. While having a common prescription at non-linear scales for all models would be desirable, given the different behaviours of each model at non-linear scales, this remains an ambitious goal. Currently, we have at our disposal codes implementing the specific features of each model that have already been tested and used in the literature. This reinforces the validity of our procedure.

\subsubsection{Jordan-Brans-Dicke gravity} \label{sec:nl-JBD}

For the non-linear prescription, we used a modified version of \texttt{HMCODE} \citep{Mead:2015yca,Mead:2016zqy}. \texttt{HMCODE} is an augmented halo model that can be used to  accurately predict  the non-linear matter power spectrum over a wide range of cosmologies. A brief summary of how this works is given below.

In the halo model \citep{Cooray:2002dia}, the non-linear matter power spectrum can be written as a sum $P_{\rm NL}(k,z) = P_{1\mathrm{H}}(k,z)+P_{2\mathrm{H}}(k,z)$ where
\begin{equation}
P_{2\mathrm{H}}(k,z)=P_{\mathrm{lin}}(k,z)
\left[\int_0^\infty b(M,z)W(M,k,z)\,n(M,z)\;\mathrm{d}M\right]^2\,,
\label{eq:two_halo}
\end{equation}
is the so-called two-halo term (correlation between different haloes), and
\begin{equation}
P_{1\mathrm{H}}(k,z)=\int_0^\infty W^2(M,k,z)\,n(M,z)\;\mathrm{d}M\,,
\label{eq:one_halo}
\end{equation}
is the so-called one-halo term (correlations between mass-elements within each halo). Above $M$ is the halo mass, $P_{\mathrm{lin}}(k,z)$ is the linear matter power spectrum, $n(M,z)$ is the halo mass function and $b(M,z)$ is the linear halo bias.

The window function $W$ is the Fourier transform of the halo matter density profiles:
\begin{equation}
W(M,k,z)=\frac{1}{\bar\rho}\int_0^{r_\mathrm{v}}4\pi r^2\frac{\sin(kr)}{kr}\,\rho(M,r,z)\;\mathrm{d}r\,,
\label{eq:window_function}
\end{equation}
where $\rho(M,r,z)$ is the radial matter density profile in a host halo with a mass $M$, and $\bar{\rho}$ is the mean matter density. The halo mass is related to the virial radius, $r_\mathrm{v}$, via
$M = 4\pi r_\mathrm{v}^3\Delta_\mathrm{v}(z)\bar\rho/3$, where $\Delta_\mathrm{v}(z)$ is the virial halo overdensity. The halo profiles, the halo definition, and the halo mass function can either be computed from excursion set models, extracted from numerical simulations and/or parametrised as functions with free parameters that are then fit to data. \texttt{HMCODE} combines this,  and the resulting fitting function has for \lcdm been shown to be accurate to $2.5\%$ for scales $k<10\, h\,{\rm Mpc}^{-1}$ and redshifts $z<2$ \citep{Mead:2020vgs}.

\cite{Joudaki:2020shz} modifed the \texttt{HMCODE}, whcih is able to include the effects of JBD. This was done using a suite of {\it N}-body simulations obtained with modified versions of both the COmoving Lagrangian Acceleration \citep[COLA,][]{Tassev:2013pn,Winther:2017jof} and \texttt{RAMSES} \citep{Teyssier:2001cp} codes. COLA solves for the perturbations around paths predicted from second-order Lagrangian perturbation theory (2LPT), and it has proven to be fast and accurate on large scales. This was used for scales $k<0.5\, h\,{\rm Mpc}^{-1}$ to generate a large enough ensemble of simulations and to reduce sample variance. On very small-scales, up to $k<10\, h\,{\rm Mpc}^{-1}$,   the \texttt{RAMSES} grid-based hydrodynamical solver with adaptive mesh refinement was used.

The spectra generated by these simulations were then used to calibrate \texttt{HMCODE}. While we did not consider the effects of baryons here, the advantage of \texttt{HMCODE}, over \texttt{Halofit} is that the former is able to capture baryonic feedback at an accuracy level of $\sim 5\%$  up to $k\simeq 10\, h\, {\rm Mpc}^{-1}$ \citep{Mead:2020vgs}. This will become an important consideration \citep{Schneider:2019snl,Schneider:2019xpf} in future more detailed analyses. To take JBD into account in addition to modifying the expansion history and growth of perturbations, the virialised halo overdensity was modified as
\begin{align}
\Delta_{\rm v} = & \Omega_{\rm m}(z)^{-0.352}\\
& \times \left\{d_0 + (418.0 - d_0) \arctan\left[(0.001|\omega_{\rm BD}-50.0|)^{0.2} \right]\frac{2}{\pi}\right\}\,,\nonumber
\end{align}
where $d_0 = 320.0 + 40.0z^{0.26}$. This modification has the feature that it reduces to the original \lcdm\ \texttt{HMCODE} as $\omega_{\rm BD}\to\infty$. We stress that this modification is not a physical claim about JBD causing this particular change in the virialised halo overdensity, but rather that this change in the virialised halo overdensity within the \texttt{HMCODE} machinery is able to accurately reproduce the JBD power spectrum to the given accuracy. The resulting power spectrum with this modification was found to be accurate to $10\%$ for the fitted range $10^4\gtrsim\omega_{\rm BD}\geq 50$ and on scales up to $k\simeq 10\, h\, {\rm Mpc}^{-1}$. This non-linear prescription was made with past weak-lensing surveys in mind and fitted to a range of $\omega_{\rm BD}$ and scales that we can realistically constrain at the present time. For a  parameter inference with actual \Euclid data the accuracy here might not be good enough.

 We show in Fig.~\ref{fig:MG-observables} the matter power spectrum and the lensing angular power spectrum of JBD with $\omega_{\rm BD}=800$, which we  refer to as JBD1, and their comparison to \lcdm.
In the top left panel we plot the linear (dashed) and non-linear (solid) matter power spectrum as a function of scale for $z=0$. 
In the top right panel we show the ratio of these power spectra with respect to their corresponding \lcdm\ cases, using the same $\sigma_8$ normalisation today.

As was shown in detail in \cite{Joudaki:2020shz}, the linear growth rate will undergo a scale-independent enhancement (without altering the shape of the linear power spectrum) while the non-linear growth will be mildly suppressed on smaller scales due to the presence of the scalar field. This is clear for JBD1, where the linear power spectrum is almost identical to its standard model counterpart, while the non-linear power spectrum shows a small suppression on small-scales that sets in at about~$k=0.3\, h\,{\rm Mpc}^{-1}$.
It is important to note that, unlike nDGP, the JBD theory is not endowed with gravitational screening, so that any modification on small-scales in the non-linear power spectrum is inherited from the change in the primordial amplitude, $A_{\rm s}$, in the linear power spectrum. In order for $\sigma_8$ to be the same in \lcdm\ as in JBD we need a higher primordial amplitude $A_{\rm s}$. A higher $A_{\rm s}$ in \lcdm\ means that non-linear structures form faster and become more massive, which in the matter power spectrum then translates into an even higher amplitude on non-linear scales (i.e. it increases more than linear theory predicts) which gives rise to the small-scale suppression seen in Fig.~\ref{fig:MG-observables}. This effect can also be found in pure \lcdm\ when we consider the ratio of the matter power spectrum for two different values of $A_{\rm s}$. The result is then constant on large scales, but shows an enhancement or suppression on small-scales. When we instead show results with the same primordial amplitude in JBD as in \lcdm\ then the JBD power spectrum would show a certain enhancement to \lcdm\ on linear scales and an ever stronger enhancement on non-linear scales, which is more in line with naive expectations of a larger gravitational constant leading to an enhancement in the matter power spectrum. 
The signature  in the matter power spectrum naturally impacts the lensing spectrum, which shows a similar suppression at small scales because for a given multipole and redshift bin, it is just proportional to the matter power spectrum as defined in Eq.~(\ref{eq:ISTrecipe}).

\subsubsection{Dvali-Gabadadze-Porrati gravity} \label{sec:nl-DGP}

We modelled the non-linear power spectrum using the halo model reaction \citep{Cataneo:2018cic} which has been shown to agree with \textit{N}-body simulations at the $2\%$ level down to $k=3 h\,{\rm Mpc}^{-1}$ with small variation depending on redshift, degree of modification to GR and mass of neutrinos \citep{Bose:2021mkz}. The approach attempts to model non-linear corrections to the power spectrum  from modified gravity through the so-called reaction $\mathcal{R}(k,z)$, which employs both one-loop perturbation theory and the halo model. The non-linear power spectrum in nDGP is given by the product 
\begin{equation}
 P_{\rm NL}(k,z) = \mathcal{R}(k,z) P^{\rm pseudo}_{\rm NL}(k,z)\,,
\end{equation}
where the \emph{pseudo}-power spectrum is a spectrum in which all non-linear physics are modelled using GR but the initial conditions are tuned in such a way as to replicate the modified linear clustering at the target redshift. We used the halofit formula of \cite{Takahashi:2012em} to model $P^{\rm pseudo}_{\rm NL}(k,z)$ by providing the  halofit formula a linear matter power spectrum modelled within nDGP gravity as input. This ensures that we have the correct modified linear clustering at $z$ while keeping a non-linear clustering as described by GR, in line with our definition of the pseudo-power spectrum.

The halo model reaction, $\mathcal{R}(k,z)$, is given by a corrected ratio of target to pseudo-halo model spectra,
\begin{equation}
\mathcal{R}(k,z) =  \frac{\{[1-\mathcal{E}(z)]e^{-k/k_\star(z)} + \mathcal{E}(z)\} P_{\rm 2H}(k,z)  +  P_{\rm 1H}(k,z)}{P_{\rm hm}^{\rm pseudo}(k,z)}\,. \label{eq:reaction}
\end{equation}
The components are given explicitly as 
\begin{align}
  P_{\rm hm}^{\rm pseudo}(k,z) = &   P_{\rm 2H} (k,z) + P_{\rm 1H}^{\rm pseudo}(k,z)\,, \label{Pk-halos} \\ 
  \mathcal{E}(z) =& \lim_{k\rightarrow 0} \frac{ P_{\rm 1H}^{\rm }(k,z)}{ P_{\rm 1H}^{\rm pseudo}(k,z)} \,, \label{mathcale} \\ 
   k_{\rm \star}(z) = & - \bar{k} \left\{\ln \left[ 
    \frac{A(\bar{k},z)}{P_{\rm 2H}(\bar{k},z)} - \mathcal{E}(z) \right] - \ln\left[1-\mathcal{E}(z) \right]\right\}^{-1}\,, \label{kstar}
\end{align}
where 
\begin{equation}
    A(k,z) =  \frac{P_{\rm 1-loop}(k,z)+ P_{\rm 1H}(k,z)}
    {P^{\rm pseudo}_{\rm 1-loop}(k,z)+ P_{\rm 1H}^{\rm pseudo}(k,z)}  P_{\rm hm}^{\rm pseudo}(k,z) -  P_{\rm 1H}(k,z)\,.
\end{equation}
$P_{\rm 2H}(k,z)$ is the two-halo contribution, which we can approximate with the (nDGP) linear power spectrum, $P_{\rm L}(k,z)$ \citep[see for a review on the halo model][]{Cooray:2002dia}. $P_{\rm 1H}(k,z)$ and $P_{\rm 1H}^{\rm pseudo} (k,z)$ are the one-halo contributions to the power spectrum as predicted by the halo model, with and without modifications to the standard \lcdm\ spherical collapse equations, respectively. We recall that by definition, the pseudo-cosmology has no non-linear beyond-\lcdm\ modifications. Similarly, $P_{\rm 1-loop}(k,z)$ and $P_{\rm 1-loop}^{\rm pseudo} (k,z)$ are the one-loop predictions with and without non-linear modifications to \lcdm, respectively. As in previous works, the limit in Eq.~(\ref{mathcale}) was taken to be at $k=0.01\,h\,{\rm Mpc}^{-1}$, and we computed $k_\star$ using $\bar{k} = 0.06\,h\,{\rm Mpc}^{-1}$. This scale was chosen such that the one-loop predictions are sufficiently accurate at all the redshifts considered.

We note that the correction to the halo-model ratio in Eq.~(\ref{eq:reaction}) has been shown to improve this ratio when there are modifications to gravity that invoke some sort of screening mechanism \citep{Cataneo:2018cic}.

As in previous works, in our halo-model calculations, we used a Sheth-Tormen mass function \citep{Sheth:1999mn,Sheth:2001dp}, a power-law concentration-mass relation \citep[see for example][]{Bullock:1999he} and an NFW halo-density profile \citep{Navarro:1996gj}. These calculations were performed using the publicly available {\tt ReACT} code \citep{Bose:2020wch}. We refer to this reference for further computational and theoretical details.

\begin{figure*}[ht!]
 \centering
 \includegraphics[width=0.85\linewidth]{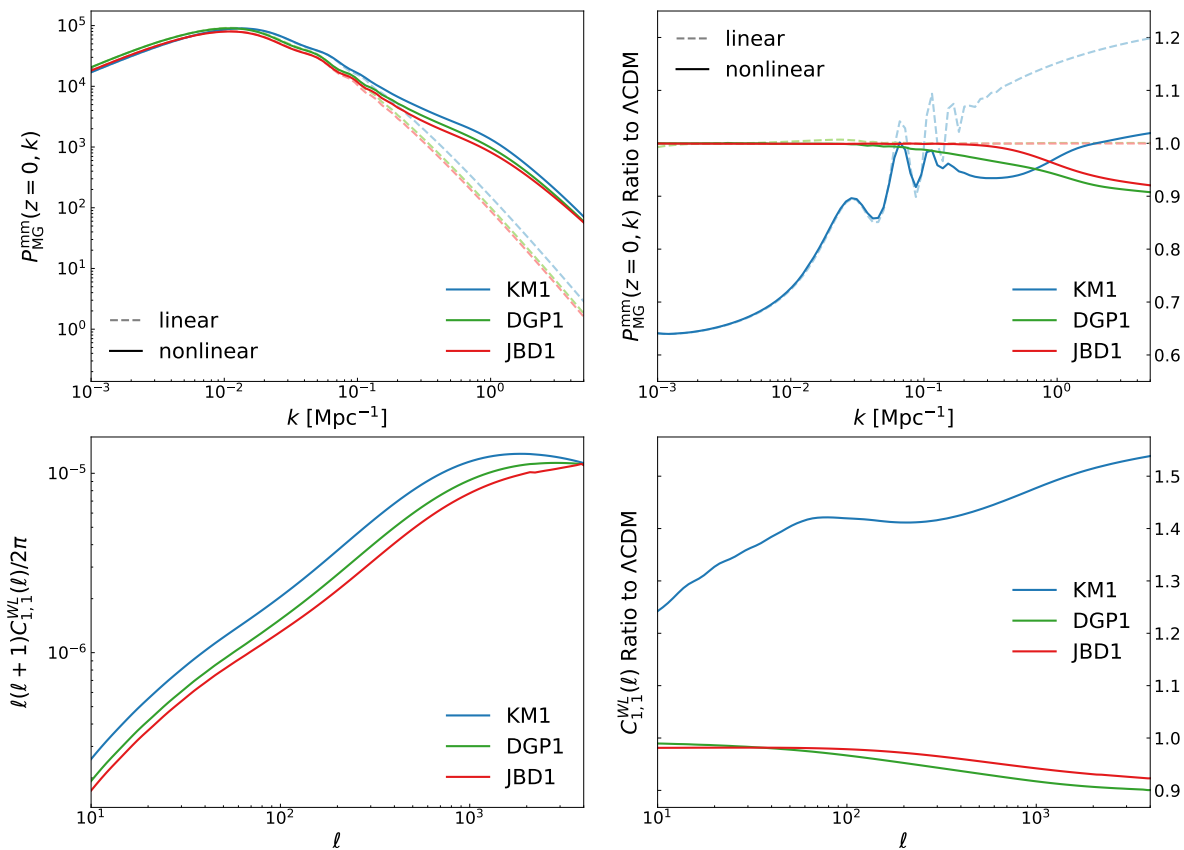}
 \caption{Large-scale structure observables for the different MG models evaluated at the fiducial values of the model parameters used for the Fisher matrix analysis.
 {\bf Top left}: Linear (dashed) and non-linear (solid lines) matter power spectrum $P_{\delta \delta}(z,k)$ entering Eqs.~(\ref{eq:pk_dw}) and (\ref{eq:ISTrecipe}), respectively, evaluated at redshift $z=0$, for KM1 ($\epsilon_{2,0}=-0.04$, blue), JBD1 ($\omega_{\rm BD}=800$, red) and nDGP1 ($\Omega_{\rm rc}=0.25$, green).
 {\bf Top right}: Ratio of the matter power spectra for the above mentioned models with respect to their \lcdm\ counterpart with the same value of $\sigma_8$ today,  for the linear (dashed) and non-linear (solid) cases. 
 {\bf Bottom left}: Cosmic shear (WL) angular power spectra for the auto-correlation of the first photometric bin $C^{\rm WL}_{1,1}(\ell)$ (solid line) and the cross-correlation of the first and last bin $C^{\rm WL}_{1,10}(\ell)$ (dotted line), defined in Eq.~(\ref{eq:ISTrecipe}) for the three models considered above, with the same colour labelling.
 {\bf Bottom right}: Ratio of the cosmic shear $C_{i,j}(\ell)$ to their \lcdm\ counterpart, for the bin combination $i=j=1$ and $i=1; j=10$. 
}
 \label{fig:MG-observables}
\end{figure*}

Finally, as previously discussed in the JBD case, we show in Fig.~(\ref{fig:MG-observables}) the comparison of the matter power spectrum and the lensing angular power spectrum of nDGP versus \lcdm\ as a function of scale for $z=0$. We chose for $\Omega_{\rm rc}$ the value $\Omega_{\rm rc}=0.25$, which  is  the largest nDGP modification  to \lcdm{}  used in this work. We  refer to this choice as the nDGP1 model. The ratio of these power spectra with respect to their corresponding \lcdm\ cases was determined using the same $\sigma_8$ normalisation today.
For nDGP1, the linear matter power spectrum is  identical to that of the \lcdm. However, the non-linear matter power spectrum shows a suppression starting from scales $k=0.1\,h\,{\rm Mpc}^{-1}$, which reaches about $10\%$ at scales of about $k=1.0\, h\,{\rm Mpc}^{-1}$. This impacts  the lensing spectrum, which shows a similar suppression at small angular scales because for a given multipole and redshift bin, it is just proportional to the matter power spectrum, as defined in Eq.~(\ref{eq:ISTrecipe}).

As the $\sigma_8$ normalisation is the same for the \lcdm\ reference spectrum, it is essentially the $P_{\rm NL}^{\rm pseudo}$ by definition. This means the quantity associated with the solid green curve in the top right panel of Fig.~\ref{fig:MG-observables} is approximately $\mathcal{R}(k,z=0)$. This quantity is largely governed by the ratio of the nDGP to pseudo one-halo terms at $k>0.1\,h\,{\rm Mpc}^{-1}$ \citep{Cataneo:2018cic}. In a nDGP cosmology, there will be more high-mass halos when compared to its GR counterpart due to the supplemental fifth force sourced by the additional degree of freedom. High-mass contributions to the one-halo term are relatively more suppressed by the NFW density profile \citep[see for example Fig.~9 of][and Eq.~\ref{Pk-1H}]{Cooray:2002dia}. This will cause the 1-halo GR power spectrum to be larger than the DGP one-halo spectrum if the linear clustering in the two cosmologies is the same at the target redshift. 

We note that when we instead  fix the same primordial amplitude of perturbations, we obtain the reverse: More halos will have formed by the target redshift in the nDGP cosmology, which means an enhancement of the one-halo term over GR. We refer  to Figs.~6 and~8 in \cite{Cataneo:2018cic}, which highlight these two scenarios, GR and nDGP, with the same late-time amplitude of linear perturbations and one with the same early-time amplitude, respectively.

\subsubsection{\texorpdfstring{$k$}{k}-mouflage gravity} \label{sec:nl-kmouflage}

For the KM non-linear matter power spectrum, we used a similar analytical approach that combined one-loop perturbation theory with a halo-model. This method was introduced in \cite{Valageas:2013gba} for the standard \lcdm\ cosmology. As in usual halo models, it splits the matter power spectrum over two-halo and one-halo contributions, as in Eq.~(\ref{Pk-halos}), but it uses a Lagrangian framework to evaluate these two contributions in terms of the pair-separation probability distribution. This provides a Lagrangian-space regularisation of perturbation theory, which matches standard perturbation theory up to one-loop order and includes a partial resummation of higher-order terms, such that the pair-separation probability distribution is positive and normalised to unity at all scales.
Thus, the two-halo term goes beyond the Zeldovich approximation by including a non-zero skewness, which enables the consistency with standard perturbation theory up to one-loop order, as well as a simple Ansatz for higher-order cumulants inspired by the adhesion model.
The one-halo term includes a counter-term that ensures its falloff at low $k$, in agreement with the conservation of matter and momentum,
\be
P_{\rm 1H}(k) = \int_0^{\infty} \frac{\de\nu}{\nu} f(\nu) \frac{M}{\bar \rho (2\pi)^3}
\left[ \tilde u_M(k) - \tilde W(k q_M) \right]^2\,,
\label{Pk-1H}
\ee
where $\tilde u_M(k)$ is the normalised Fourier transform of the halo radial profile, $\tilde W(kq_M)$ is the normalised Fourier transform of the top-hat of Lagrangian radius $q_M$, and $f(\nu)$ is the normalised halo mass function. 
This counter-term is not introduced by hand, as it arises directly from the splitting of the matter power spectrum over two-halo and one-halo contributions within a Lagrangian framework, which by construction satisfies the conservation of matter. Additionally, $\nu$ is defined as $\delta_c/\sigma(M)$, where $\delta_{\rm c}$ is the linear density contrast associated with a non-linear density contrast of 200, and $\sigma(M)$ is the root mean square of the linear density contrast at mass scale $M$. The threshold $\delta_{\rm c}$ is sensitive to the modification of gravity. In principle, it also depends on $M$, as the shells in the spherical collapse are coupled and evolve differently depending on their masses due to non-linear screening. However, for the models that we considered here, screening is negligible beyond $1\,h^{-1}$ Mpc (clusters are not screened), so that $\delta_{\rm c}$ is independent of $M$.

This non-linear modelling was extended in \cite{Brax:2013fna} to several modified-gravity scenarios [$f(R)$ theories, dilaton, and symmetron models], and in \cite{Brax:2014yla,Brax:2015lra} to KM models.
This involved the computation of the impact of modified gravity on both the linear and one-loop contributions to the matter power spectrum, which enter the two-halo term, and on the spherical collapse dynamics, which enter the one-halo term through the halo mass function.
Therefore, this non-linear modelling exactly captures the modification of gravity both up to one-loop order and at the level of the fully non-linear spherical collapse.
The Ansatz for the partial resummation of higher-order terms is the same as for the standard \lcdm\ cosmology (this implies that their values are also modified as they depend on the lower orders to ensure the positivity and the normalisation of the underlying pair distribution).
A comparison with numerical simulations \citep{Brax:2013fna} for $f(R)$ theories (with $|f_{R_0}| = 10^{-4}$, $10^{-5}$ and $10^{-6}$), shows that this approach captures the relative deviation from the \lcdm\ power spectrum at $z=0$ up to $k=3\, h \, {\rm Mpc}^{-1}$.
We have not yet compared this recipe with numerical simulations of the KM theories, which have only recently been performed \citep{Hernandez-Aguayo:2021kuh}. However, we expect at least the same level of agreement because the KM models are simpler and closer to the \lcdm\ cosmology than the $f(R)$ models (as in \lcdm, the linear growth rate is scale independent).

 Fig.~\ref{fig:MG-observables} shows the comparison between the KM1 model defined with $\epsilon_{2,0}=-0.04$ and its \lcdm\ counterpart, normalised to the same $\sigma_8$ and cosmological parameters at $z=0$, for the matter and cosmic shear power spectra. The ratio of the linear matter power spectra increases on intermediate scales ($10^{-2}<k< 1 \, h\,{\rm Mpc}^{-1}$): This is due to the change of the background evolution, which slightly affects the scales at matter-radiation equality.
$H(z)$ is indeed greater for KM1 at high redshift \citep{Benevento_2019}, which implies a smaller Hubble radius at matter-radiation equality. This in turns means that the change in the slope of the linear power spectrum, from $P_{\rm L}(k) \sim k^{n_{\rm s}}$ to $P_{\rm L}(k) \propto k^{n_{\rm s}-4}$, is shifted to smaller scales, and  hence to higher $k$. This implies that the ratio $P^{\rm mm}_{\rm KM1}/P^{\rm mm}_{\lcdm}$ grows on these intermediate scales, where the linear power spectrum changes slope, because of this delay to higher $k$ of the transition. 
Because the two spectra have the same $\sigma_8$ and the ratio $P^{\rm mm}_{\rm KM1}/P^{\rm mm}_{\lcdm}$ grows with $k$, the ratio is below unity at low $k$ and above unity at high $k$.
Moreover the linear growth rates are scale independent for both KM and \lcdm, and therefore the ratio $P^{\rm mm}_{\rm KM1}/P^{\rm mm}_{\lcdm}$ reaches a constant at very low $k$ and at very high $k$, where the slopes of both linear power spectra are equal to $n_{\rm s}$ or $n_{\rm s}-4$ (i.e. outside of the intermediate scales where the slope of the linear power spectrum slowly runs with $k$).
This flat ratio is clearly shown in Fig.~\ref{fig:MG-observables} at low $k$.  The amplitude of $P^{\rm mm}_{\rm KM1}/P^{\rm mm}_{\lcdm}$ is set by the parameter $\epsilon_{2,0}$, and it deviates substantially from 1, by about $30\%$, for a value of $\epsilon_{2,0} = -0.04$.
The increase in $P^{\rm mm}_{\rm KM1}/P^{\rm mm}_{\lcdm}$ with $k$ is partly due to non-linear effects in the scalar field sector, both in the background and in the linear growth rate, which are not completely irrelevant. At $z=0$, even though $\epsilon_2=-0.04$ we have $\epsilon_1 \simeq 0.12$, whereas at linear order in the scalar field dynamics we would have expected that $\epsilon_1=-\epsilon_2$. As the growth of structures is  sensitive to $\epsilon_1$, this leads to a stronger effect than could be expected from the value of $\epsilon_2$ itself. Then, the power spectrum being a quadratic functional of the density field gives an additional amplification factor of $2$ for relative perturbations.

Beyond this global behaviour of the ratio $P^{\rm mm}_{\rm KM1}/P^{\rm mm}_{\lcdm}$, another impact of the slight modification of the background is that the baryon acoustic oscillations are shifted. This leads to the oscillations seen at BAO scales, $k \sim 0.01\, h\,{\rm Mpc}^{-1}$.
Non-linear effects lead to a decrease in the ratio $P^{\rm mm}_{\rm KM1}/P^{\rm mm}_{\lcdm}$ as compared with the linear prediction of mildly non-linear scales. This is due to the loss of information in the power spectrum in the non-linear regime, which can be related to the universal NFW profile of virialised haloes within the halo model. Then, the shape of the non-linear power spectrum at those scales is mostly set by the scale where the non-linear transition takes place, and this occurs to occur at somewhat larger scales for the \lcdm\ counterpart (i.e. the higher linear power at higher $k$ in the KM1 scenario is mostly erased by the non-linear dynamics).

As usual, the integration along the line of sight leads to a much smoother curve for the cosmic shear angular power spectrum. Even though the 3D non-linear matter power spectrum is reduced at $z=0$ for KM1 as compared with \lcdm, as shown in the upper right panel, the lensing power is increased, as shown in the lower right panel. This is due to the coupling function $A(\varphi)$, which is greater than unity at $z>0$ and leads to a higher value of the phenomenological function $\Sigma$, see Eq.~(\ref{mu_Sigma_KM}), or in other words, to a greater Newton constant at high redshift.

\section{Survey specifications and analysis method} \label{sec:fisher}

We varied as base cosmological parameters:
\begin{equation}
  \bm\Theta=\{\Omega_{\rm m,0},\, \Omega_{\rm b,0},\, h,\, n_{\rm s},\, \sigma_8\}\,,
\end{equation}
for which we chose the following fiducial values:
\begin{align}
 \Omega_{\rm m,0} & = 0.315\,,
 & \Omega_{\rm b,0} & = 0.049\,, \nonumber\\
 h & = 0.6737\,, & n_{\rm s} & = 0.966\,,
\end{align}
where $\Omega_{\rm m,0}= \Omega_{\rm c,0}+\Omega_{\rm b,0}+\Omega_{\rm \nu,0}$.
To fix the fiducial value for $\sigma_8$, we  used the same initial amplitude of primordial perturbations for all models, namely $A_{\rm s} = 2.09681 \times 10^{-9}$.
The fiducial cosmology includes massive neutrinos with a total mass of $\Sigma m_\nu=0.06\,\mathrm{eV}$, but we kept $\Sigma m_\nu$ fixed in the following Fisher matrix analysis. 
In the following, we  specify the $\sigma_8$ value for each model along with the fiducial values of the model parameters,

\begin{enumerate}
 \item JBD
  \begin{align}
   \bm\Theta_{\rm fid,1}&=\{ \sigma_8=0.816,\, \log_{10}\omega_{\rm BD} = 2.90309 \}\, \quad({\rm JBD1});\nonumber \,\\
   \bm\Theta_{\rm fid,2}&=\{  \sigma_8=0.812,\, \log_{10}\omega_{\rm BD}= 3.39794\}\, \quad({\rm JBD2}).
   \label{eq:fiducial-jbd}
  \end{align}
  These values correspond to $\omega_{\rm BD}= 800$ and $\omega_{\rm BD}= 2500$ respectively. In these cases the initial value of the scalar field was fixed to $\phi_{\rm ini} = 1/G_{\rm N}$.\\
 \item nDGP
  \begin{align}
   \bm\Theta_{\rm fid,1}&=\{ \sigma_8 = 0.8690,\, \log_{10}\Omega_{\rm rc} = -0.60206\}\, \quad({\rm nDGP1}); \nonumber \,\\
   \bm\Theta_{\rm fid,2}&=\{ \sigma_8 = 0.8105,\, \log_{10}\Omega_{\rm rc} = -6\}\, \quad({\rm nDGP2}).
   \label{eq:fiducial-ndgp}
  \end{align}
  These values correspond to $\Omega_{\rm rc} = 0.25$ and $\Omega_{\rm rc} = 10^{-6}$,  for nDGP1 and nDGP2, respectively.\\
 \item KM
  \begin{align}
   \bm\Theta_{\rm fid,1}&=\{ \sigma_8=0.994, \, \epsilon_2= -0.04 \}\, \quad({\rm KM1});\nonumber \,\\
   \bm\Theta_{\rm fid,2}&=\{ \sigma_8=0.813,\, \epsilon_2= -0.0001 \}\, \quad({\rm KM2}).
   \label{eq:fiducial-kmouflage}
  \end{align}
\end{enumerate}

We followed \citetalias{Euclid:2019clj} to set up the specifics of the photometric probes as we briefly summarise here. The sources were split into ten equi-populated redshift bins whose limits were obtained from the redshift distribution
\begin{equation}
    n(z)\propto\left(\frac{z}{z_0}\right)^2\,\text{exp}\left[-\left(\frac{z}{z_0}\right)^{3/2}\right]\,,
\end{equation}
with $z_0=0.9/\sqrt{2}$ and the normalisation set by the requirement that the surface density of galaxies is $\bar{n}_g=30\,\mathrm{arcmin}^{-2}$. This was then convolved with the sum of two Gaussians to account for the effect of photometric redshift. Galaxy bias was assumed to be constant within each redshfit bin, with fiducial values $b_i = \sqrt{1 + \bar{z}_i}$, and $\bar{z}_i$ the bin centre. Any possible scale dependence of the bias introduced by GR was taken to be negligible.

As in \citetalias{Euclid:2019clj}, we considered a Gaussian-only covariance, whose elements are given by
\begin{multline}
    \text{Cov}\left[C_{ij}^{\rm AB}(\ell),C_{kl}^{\rm CD}(\ell')\right]=\frac{\delta_{\ell\ell'}^{\rm K}}{(2\ell+1)f_{\rm sky}\Delta \ell}\\
    \times\left\{\left[C_{ik}^{\rm AC}(\ell)+N_{ik}^{\rm AC}(\ell)\right]\left[C_{jl}^{\rm BD}(\ell')+N_{jl}^{\rm BD}(\ell')\right]\right.\\
    +\left.\left[C_{il}^{\rm AD}(\ell)+N_{il}^{\rm AD}(\ell)\right]\left[C_{jk}^{\rm BC}(\ell')+N_{jk}^{\rm BC}(\ell')\right]\right\}\,,
\end{multline}
where the upper-case (lower-case) Latin indexes run over WL, \GCph\ (all tomographic bins), $\delta_{\ell\ell'}^{\rm K}$ is the Kronecker delta symbol coming from the lack of correlation between different multipoles $(\ell, \ell')$, $f_{\rm sky}\simeq 0.36$ is the survey sky fraction, and $\Delta \ell$ denotes the width of the $100$ logarithmic equi-spaced multipole bins. For the observables of interest here, we assumed a white noise, so that it is
\begin{align}
    N_{ij}^{\rm LL}(\ell) &= \frac{\delta_{ij}^{\rm K}}{\bar{n}_i}\sigma_\epsilon^2\,,\\
    N_{ij}^{\rm GG}(\ell) &= \frac{\delta_{ij}^{\rm K}}{\bar{n}_i}\,,\\
    N_{ij}^{\rm GL}(\ell) &= 0\,,
\end{align}
where $\sigma_\epsilon^2=0.3^2$ is the variance of the observed ellipticities.

We continued to follow \citetalias{Euclid:2019clj} to evaluate the Fisher matrix $F_{\alpha\beta}(z_i)$ for the observed galaxy power spectrum. Here, $\alpha$ and $\beta$ run over the cosmological parameters of the set $\bm\Theta$, and the index $i$ labels the redshift bin, each  centred in $z_i = \{1.0,\,1.2,\,1.4,\,1.65\}$, whose widths were $\Delta z = 0.2$ for the first three bins and $\Delta z = 0.3$ for the last bin. As a difference with respect to \citetalias{Euclid:2019clj}, we adopted  the direct derivative approach and directly varied the observed galaxy power spectrum with respect to the cosmological parameters. We included two additional redshift-dependent parameters $\ln b\sigma_8(z_i)$ and $P_{\rm s}(z_i)$ over which we marginalised. the spectroscopic galaxy bias, $b(z)$, and the expected number density of H$\alpha$ emitters, $n(z)$, are as reported in Table 3 of \citetalias{Euclid:2019clj}.

For all probes, we considered an optimistic and a pessimistic scenario defined according to the specifications for WL, \GCph\ and \GCsp\ in Table~\ref{tab:specifications-ec-survey}.  For \GCsp, we  added a third scenario, referred to as {\it quasi-linear}, for which we fixed the maximum wavenumber to $k_{\rm max} = 0.15\,h\,{\rm Mpc}^{-1}$. We explored this more conservative case since the underlying matter power spectrum of Eq.~(\ref{eq:pk_dw}) that we used in our observed galaxy power spectrum recipe is a linear one. Non-linear corrections begin to play a role for wave numbers larger than $k = 0.1\,h\,{\rm Mpc}^{-1}$ for the redshifts we considered \citep[see][]{Taruya:2010mx}. Cutting at a lower $k_{\rm max}$ should avoid any bias induced by neglecting non-linear corrections below this scale. However, this severe cut also removes part of the information encoded in \GCsp\ so that we wished to quantify how this affects the constraints. In all scenarios, we fixed the $\sigma_{\rm v}$ = $\sigma_{\rm p}$ nuisance parameter for \GCsp\ to the values directly calculated from Eq.~(\ref{eq:sigmav}) for the fiducial cosmological parameters.

\begin{table}
	\centering
	\caption{\Euclid survey specifications for WL, \GCph\ and \GCsp.}
	\label{tab:specifications-ec-survey}
	\begin{tabularx}{\columnwidth}{Xll}
		\hline 
		Survey area & $A_{\rm survey}$  & $15\,000\,\deg^2$  \\
		\hline
		\hline
		\multicolumn{3}{c}{WL}\\
		\hline
		Number of photo-$z$ bins & $N_z$ & 10 \\
		Galaxy number density & $\bar n_{\rm gal}$  & $30\,\mathrm{arcmin}^{-2}$ \\
		Intrinsic ellipticity dispersion & $\sigma_\epsilon$  & 0.30 \\
		Minimum multipole & $\ell_{\rm min}$ & 10\\
		Maximum multipole & $\ell_{\rm max}$ & \\
		\phantom{Maximum}-- Pessimistic & & $1500$\\
		\phantom{Maximum}-- Optimistic & & $5000$\\
        \hline
		\hline
        \multicolumn{3}{c}{\GCph}\\
        \hline
		Number of photo-$z$ bins & $N_z$ & 10 \\
		Galaxy number density & $\bar n_{\rm gal}$  & $30\,\mathrm{arcmin}^{-2}$ \\
		Minimum multipole & $\ell_{\rm min}$ & 10 \\
		Maximum multipole & $\ell_{\rm max}$ & \\
		\phantom{Maximum}-- Pessimistic & & $750$\\
		\phantom{Maximum}-- Optimistic & & $3000$\\
		\hline
		\hline
		\multicolumn{3}{c}{\GCsp}\\
		\hline
		Number of spectro-$z$ bins & $n_z$ & 4 \\
		Centres of the bins &$z_i$ & $\{1.0,\,1.2,\,1.4,\,1.65\}$\\
		Error on redshift & $\sigma_{0,z}$ & 0.001 \\
		Minimum scale & $k_{\rm min}$ & $0.001\,h\,{\rm Mpc}^{-1}$\\
		Maximum scale & $k_{\rm max}$ & \\
		\phantom{Maximum}-- Quasi-linear & & $0.15\,h\,{\rm Mpc}^{-1}$\\
		\phantom{Maximum}-- Pessimistic & & $0.25\,h\,{\rm Mpc}^{-1}$\\
		\phantom{Maximum}-- Optimistic & & $0.30\,h\,{\rm Mpc}^{-1}$\\
		\hline 
	\end{tabularx}
\end{table}

We considered both \GCsp\ alone and the combination of all photometric probes: \GCph, WL, and their XC. Moreover, we also considered the full combination of all \Euclid main probes: \GCsp, \GCph, WL, and the XC between \GCph\ and WL. It is important to mention that we accounted for all the mixed terms in the covariance matrix for the photometric probes such as, for instance, $\text{Cov}\left[C_{ij}^{\rm LL}(\ell),C_{kl}^{\rm GL}(\ell)\right]$. However, we followed \citetalias{Euclid:2019clj} in neglecting any correlation between \GCsp\ and the photometric probes. In the pessimistic scenario, we further imposed a cut $z < 0.9$ to \GCph\ and XC to remove any overlap between these probes and \GCsp. This cut was applied only when we added \GCsp\ to the photometric probes, while was not applied when we used the photometric probes alone in the pessimistic case. No cut was ever applied in the optimistic case.

\section{Results} \label{sec:results}
In this section, we discuss the results of the Fisher matrix analysis for the three models under investigation, JBD, nDGP, and KM, and for each of them we considered the two fiducial cases presented in Sect.~\ref{sec:fisher}. Additionally, for each case we considered a quasi-linear, pessimistic, and optimistic scenario as discussed in Sect.~\ref{sec:fisher}. In the following we present the 68.3\% and 95.4\% joint marginal error contours on the cosmological and model parameters for the optimistic setting, and for completeness, we include the contour plots for the pessimistic scenario in  Appendix~\ref{sec:appendix}. We stress that we are interested in detecting a non-zero value for the model parameters, and therefore we present the relative errors on the corresponding fiducial values. In the specific cases of JBD and nDGP, we performed the Fisher analysis on the logarithm of the additionalparameter, namely $\log_{10}X$ with $X=\omega_{\rm BD}$ and $X=\Omega_{\rm rc}$ for JBD and nDGP, respectively. This choice was dictated by the fact that we cannot perform a Fisher analysis on a parameter that varies  by some orders of magnitude. Additionally, we need to use as parameter a quantity that is of order one in the analysis in order to avoid  large differences between the highest and lowest eigenvalues of the Fisher matrix. For the specific case of nDGP, the use of the logarithm of the model parameter allowed us to consider only its positive viable range, which would not be possible if we performed a Fisher analysis directly on $\Omega_{\rm rc}$. We note indeed that a prior such as $\Omega_{\rm rc}>0$ cannot be imposed a priori in the Fisher analysis.
In addition to discussing the results on $\log_{10}X$ we  also report on the uncertainties on the parameter itself, $X$. We note that we cannot use a Jacobian transformation to convert between the Fisher matrices because the transformation between the two parametrisations is non-linear and the assumption of Gaussianity is only valid for the logarithmic parametrisation. In order to obtain the constraints on the parameter $X$, we used
\begin{equation}
\label{Eq:logXtoX}
X^{(\pm)}=X_{\rm fid}\times 10^{\pm \sigma_{\log_{10}X}}\,.
\end{equation} 
As a consequence, the uncertainties on $X$ are asymmetrical. 
We now proceed to discuss  each model in detail.

\begin{figure}[h!]
 \centering
 \includegraphics[width=
\linewidth]{barplot_scales_spec_hbug.pdf}\\
\vspace{-20pt}
\includegraphics[width=
\linewidth]{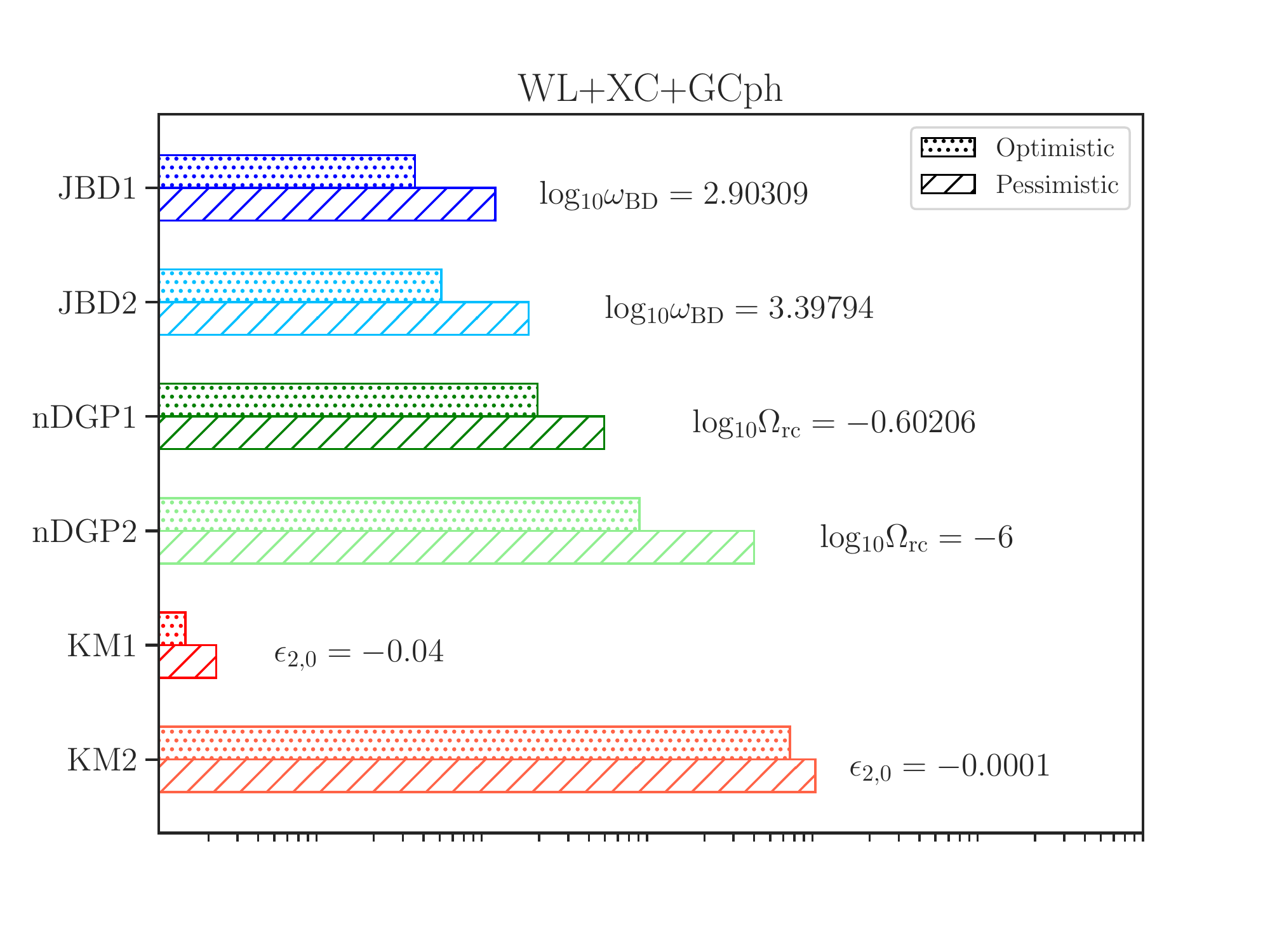}\\
\vspace{-20pt}
\includegraphics[width=
\linewidth]{barplot_scales_all_hbug.pdf}
 \caption{Marginalised $1\sigma$ errors on the model parameters relative to their corresponding fiducial values, for JBD (case 1 in blue and case 2 in cyan), nDGP (case 1 in green and case 2 in light green), and KM (case 1 in red and case 2 in light red). We show the marginalised $1\sigma$ errors for \GCsp\ in the optimistic (dotted area), pessimistic (dashed area) and quasi-linear (filled area) scenarios (top panel), for WL+XC+\GCph\ in the optimistic and pessimistic cases (central panel) and for the \GCsp+WL+XC+\GCph\ in the the optimistic and pessimistic cases.}
 \label{fig:barplot}
\end{figure}

In order to have a global view on the results for the three MG models analysed, we summarise in Fig.~\ref{fig:barplot} the 68.3\%  marginal errors on the model parameters, $\log_{10}\omega_{\rm BD}$ for JBD, $\log_{10}\Omega_{\rm rc}$ for nDGP, and $\epsilon_{2,0}$ for KM in all the scenarios and combinations of probes we used. We note that we did not consider WL probe alone because we verified that WL provides constraints of about  the same order as the \GCsp\ ones. When we combined this with \GCph\ and XC, the constraints were significantly boosted. Additionally, we note that the  constraining power of the spectroscopic sample is very weak because we did not go deep enough ($k\gtrsim 1\,h\,{\rm Mpc}^{-1}$) in the non-linear regime and also because when we used  GC probes, we introduced degeneracies among the MG parameters, amplitudes, and bias parameters. These features characterise all the models.

\subsection{Jordan-Brans-Dicke gravity}\label{Sec:ResultsnJBD}

\begin{figure*}[ht!]
 \centering
 \includegraphics[width=0.49\linewidth]{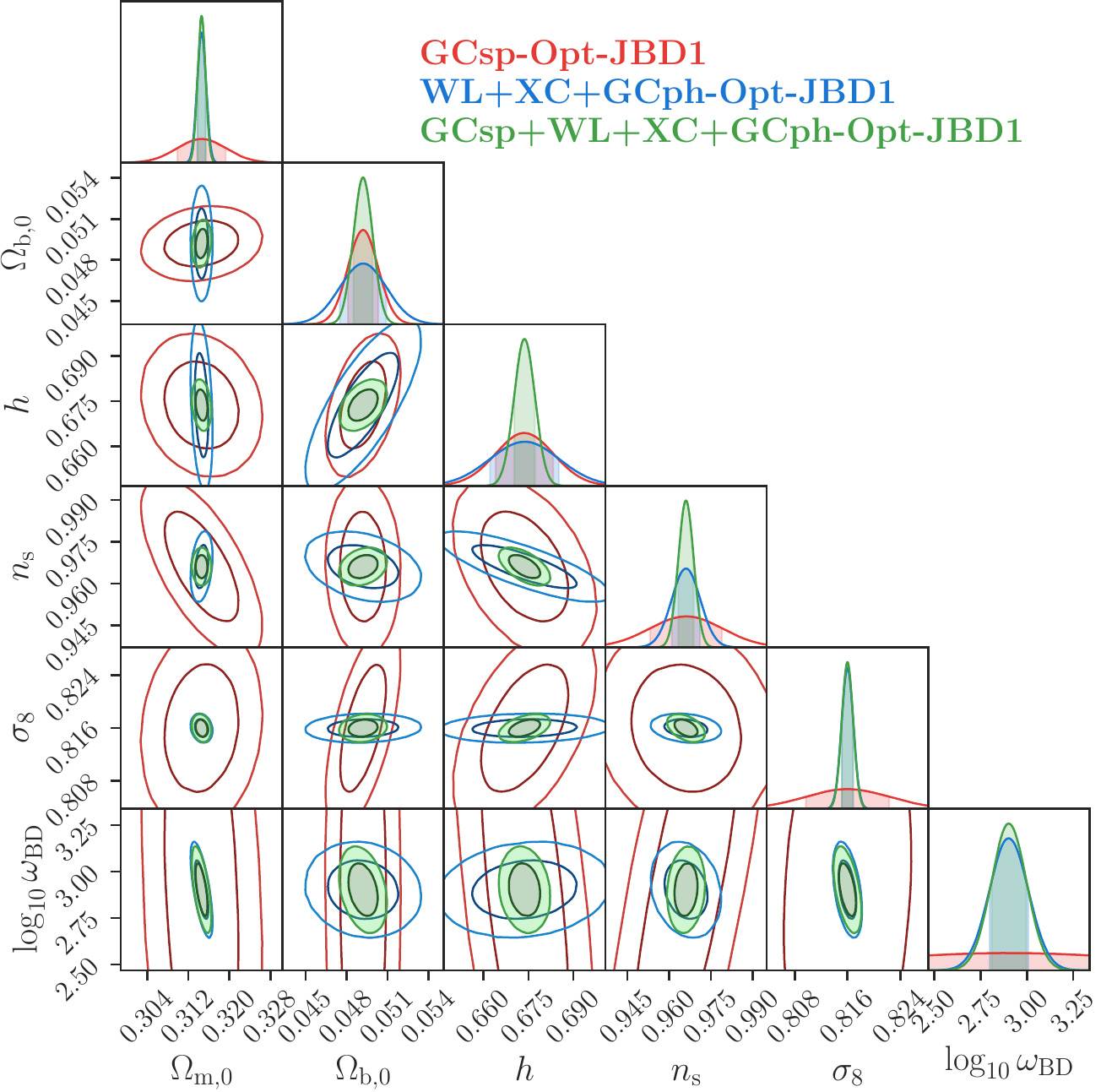}
 \includegraphics[width=0.49\linewidth]{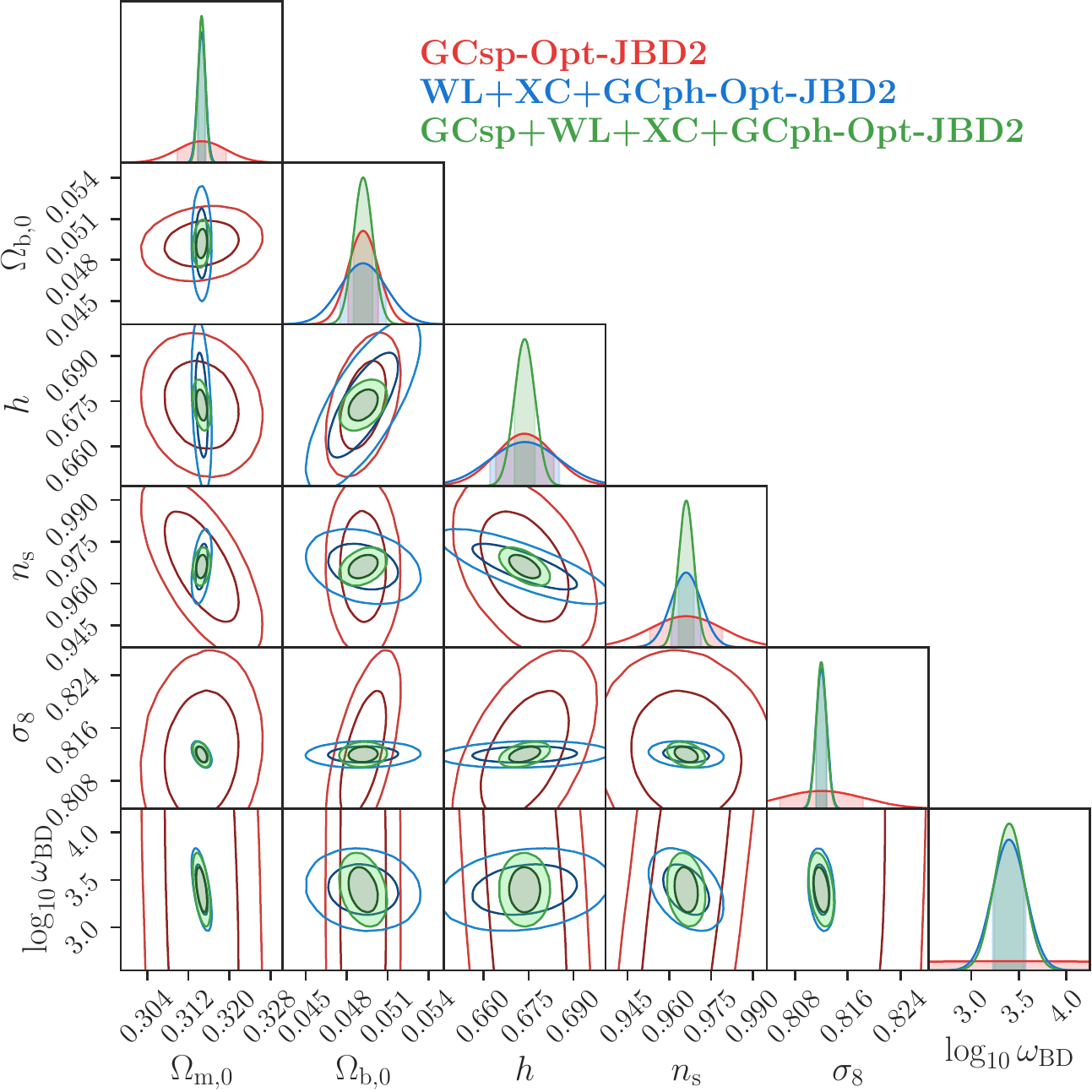}
 \caption{68.3\% and 95.4\% joint marginal error contours on the cosmological parameters for JBD1 (left panel) and JBD2 (right panel) in the optimistic case. In red, we plot \GCsp, in blue, we plot WL+XC+\GCph\ and in green we plot \GCsp+WL+XC+\GCph.
 }
 \label{fig:ellipses-JBD}
\end{figure*}

There are remarkably strong non-cosmological constraints on JBD gravity, with current bounds of about $\omega_{\rm BD}>10^5$. As pointed out above, however, it is useful to consider JBD gravity as a long-wavelength limit of more general scalar-tensor theories with a wide range of small-scale limits that may be endowed with gravitational screening. It then makes sense to focus on what we already know from cosmology, where the constraints are about $\omega_{\rm BD}>10^3$, independent of what we know from non-cosmological systems. Two further practical reasons led us to choose $\omega_{\rm BD}\sim 10^3$  as a fiducial value. First of all, this value is at the limits of the range for which we calibrated the non-linear prescription we used. Second, the likelihood is very flat for high values of $\omega_{\rm BD}$ which can lead to numerical errors in the finite differences of the Fisher matrix evaluation for some of the observables. We thus considered two cases for JBD. The first fiducial case with $\omega_{\rm BD}=800$ (JBD1), and the second case with $\omega_{\rm BD}=2500$ (JBD2). Whereas the first case is nearly compatible with the most recent constraints from publicly available CMB and LSS data, the second case is well within the current bounds \citep{Ballardini:2020iws,Joudaki:2020shz,Ballardini:2021evv}.
Table~\ref{tab:rel-errors-JBD} summarises the relative (with respect to the fiducial values) uncertainties for the cosmological and model parameters in the quasi-linear, pessimistic and optimistic \Euclid cases. 
Fig.~\ref{fig:ellipses-JBD} shows the 68.3\% and 95.4\% joint marginal error contours on the cosmological model parameters for the optimistic settings for both JBD1 (left panel) and JBD2 (right panel). As in \cite{Casas:2023lnp} for $f(R)$, we performed the Fisher matrix analysis on the parameter $\log_{10}\omega_{\rm BD}$, and not directly on $\omega_{\rm BD}$, because for large numbers and differences in the order of magnitude, the Fisher matrix derivatives might become unstable.

In the optimistic setting, we find that \Euclid will be able to constrain $\log_{10}\omega_{\rm BD}$ as follows:
\begin{itemize}
 \item JBD1:
 \begin{itemize}
     \item at 27\% for the \GCsp\ alone;
     \item at 3.6\% when considering the combination WL+XC+\GCph;
     \item at 3.2\% for \GCsp+WL+XC+\GCph.
 \end{itemize}
 \item JBD2: \begin{itemize}
      \item at 72.2\% for the \GCsp\ alone;
     \item  at 5.1\% when considering the combination WL+XC+\GCph;
     \item at 4.6\% for \GCsp+WL+XC+\GCph.
      \end{itemize}
\end{itemize}

For JBD1 the parameter $\log_{10}\omega_{\rm BD}$ can be measured at high statistical significance even in the pessimistic case, but this does not hold for JBD2.

We now discuss the forecast errors on the parameter $\omega_{\rm BD}$ propagating the uncertainties according to Eq.~(\ref{Eq:logXtoX}). 
For JBD1 we obtain $\omega_{\rm BD}= 800^{+4100}_{-670}$ for \GCsp, $\omega_{\rm BD}=800^{+210}_{-170}$ for WL+XC+\GCph\ and $\omega_{\rm BD}=800^{+200}_{-160}$ for \GCsp+WL+XC+\GCph; these results correspond to relative errors of $513\%$, $27\%$ and $24\%$, respectively.
For JBD2, always in the case of optimistic settings, we obtain $\omega_{\rm BD}=2500^{+1200}_{-820}$ for WL+XC+\GCph\ and $\omega_{\rm BD}=2500^{+1070}_{-750}$ for \GCsp+WL+XC+\GCph. This corresponds to relative errors of $49\%$ and $43\%$, respectively (JBD2 is unconstrained by GC alone). 

These results for the optimistic settings demonstrate that \Euclid alone could potentially detect at statistically significant level values of $\omega_{\rm BD}$ which are compatible with current publicly available CMB and LSS data \citep{Ballardini:2020iws,Joudaki:2020shz,Ballardini:2021evv}, but not those that are compatible with Solar System constraints. 
Although the \Euclid specifications we used here are different from those used in \cite{Ballardini:2019tho}, the results we find are consistent with theirs.

\begin{table*}[htbp]
\caption{Forecast $1\sigma$ marginal errors on the cosmological and model parameters relative to their corresponding fiducial value for JBD1 and JBD2 in the pessimistic, quasi-linear and optimistic cases, using \Euclid observations \GCsp, WL+XC+\GCph\ and \GCsp+WL+XC+\GCph.
}
\begin{tabularx}{\textwidth}{Xcccccc}
\hline
\rowcolor{jonquil} \multicolumn{7} {c}{{{\mbox{\textbf{JBD1}}\,\,\, \boldsymbol{$\omega_{\rm BD}=800$}}}} \\
& \multicolumn{1}{c}{$\Omega_{\rm m,0}$} & \multicolumn{1}{c}{$\Omega_{\rm b,0}$} & \multicolumn{1}{c}{$\log_{10}\omega_{\rm BD}$} & \multicolumn{1}{c}{$h$} & \multicolumn{1}{c}{$n_{\rm s}$} & \multicolumn{1}{c}{$\sigma_{8}$} \\
\hline
\rowcolor{lavender(web)}\multicolumn{7}{l}{{{Pessimistic setting}}} \\ 
\GCsp\ $(k_{\rm max} = 0.15\,h\,{\rm Mpc}^{-1})$ (quasi-linear) & 2.13\%	& 4.49\% & 43.04\%	&	1.72\%	&	2.62\%	&	1.11\%  \\
GC$_{\rm sp}\,(k_{\rm max} = 0.25\,h\,{\rm Mpc}^{-1})$ &	1.55\%	&	2.48\%	&	29.62\%	&	1.72\%	&	1.53\%	&	0.83\%  \\
WL+XC+\GCph & 0.37\% & 5.10\% &	10.89\%	& 3.20\% & 1.25\% &	0.34\% \\
\GCsp+WL+XC+\GCph &	0.36\%	&	1.72\%	&	9.21\%	&	0.91\%	&	0.66\%	&	0.28\%  \\
\hline
\rowcolor{lavender(web)}\multicolumn{7}{l}{{{Optimistic setting}}}  \\ 
\GCsp\ $(k_{\rm max} = 0.3\,h\,{\rm Mpc}^{-1})$  &	1.49\%	&	1.21\%	&	27.13\%	&	1.41\%	&	1.33\%	&	0.77\%  \\
WL+XC+\GCph &	0.27\%	&	3.43\%	&	3.55\%	&	1.69\%	&	0.52\%	&	0.11\% \\
\GCsp+WL+XC+\GCph & 0.24\%	& 1.43\% & 3.21\% &	0.51\% & 0.28\%	& 0.10\%  \\
\hline
\hline
\rowcolor{jonquil} \multicolumn{7} {c} {{{\mbox{\textbf{JBD2}}\,\,\, \boldsymbol{$\omega_{\rm BD}=2500$}}}} \\ 
\hline
\rowcolor{lavender(web)}\multicolumn{7}{l}{{{Pessimistic setting}}} \\
\GCsp\ $(k_{\rm max} = 0.15\,h\,{\rm Mpc}^{-1})$ (quasi-linear)  & 2.14\% & 4.52\%	& 114.54\% & 3.80\% & 2.64\% & 1.11\%  \\
\GCsp\,$(k_{\rm max} = 0.25\,h\,{\rm Mpc}^{-1})$ & 1.55\% &	2.49\%	& 78.82\% & 1.74\% & 1.53\% & 0.83\%  \\
WL+XC+\GCph & 0.34\% & 5.09\% & 17.32\%	& 3.21\% & 1.24\% & 0.27\% \\
\GCsp+WL+XC+\GCph &	0.33\%	& 1.68\% &	15.51\% & 0.78\% & 0.54\% & 0.24\%  \\
\hline
\rowcolor{lavender(web)}\multicolumn{7}{l}{{{Optimistic setting}}}  \\ 
\GCsp\ $(k_{\rm max} = 0.3\,h\,{\rm Mpc}^{-1})$ & 1.50\% & 2.24\% &	72.20\% & 1.42\% & 1.34\% & 0.77\%  \\
WL+XC+\GCph & 0.25\% &	3.40\%	& 5.12\% & 1.70\% & 0.56\% & 0.10\% \\
\GCsp+WL+XC+\GCph &	0.22\%	& 1.42\% &	4.56\%	& 0.50\% &	0.28\% & 0.09\%  \\
\hline
\hline
\end{tabularx}
\label{tab:rel-errors-JBD}
\end{table*}

\subsection{Dvali-Gabadadze-Porrati gravity}\label{Sec:ResultsnDGP}

\begin{figure*}[ht!]
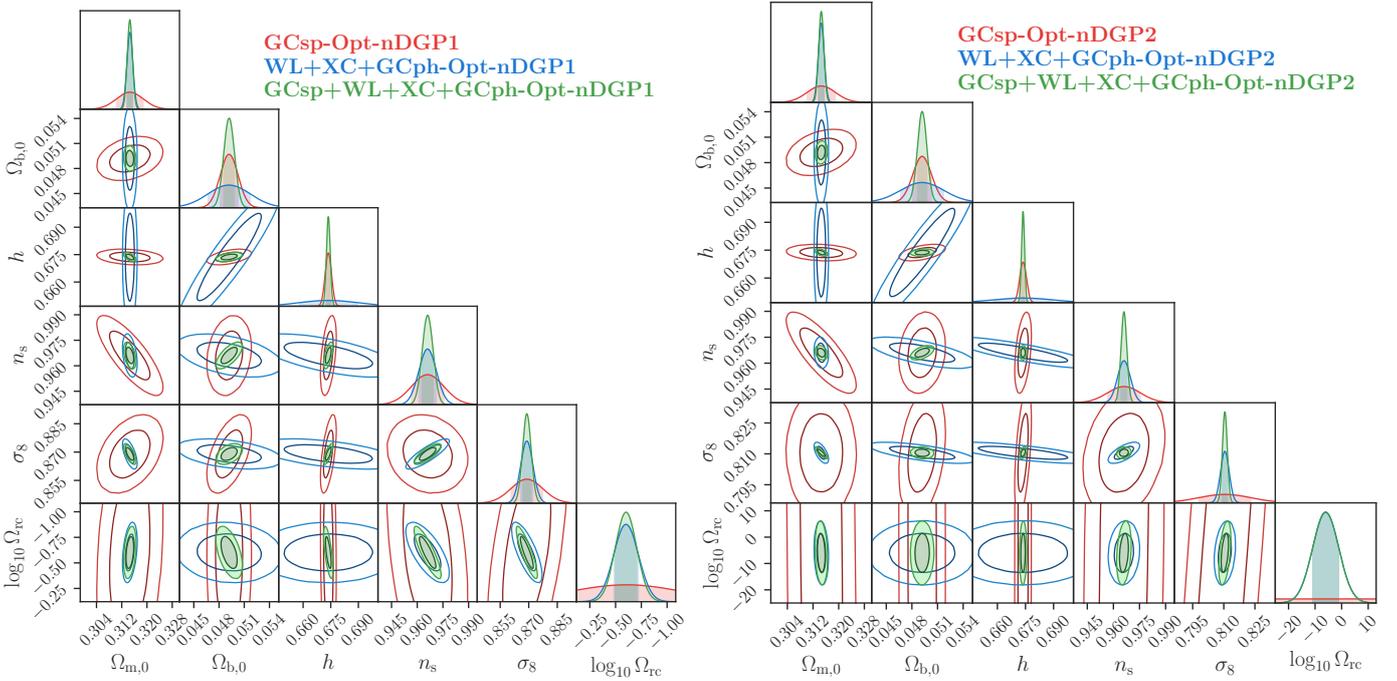

 \centering
 \includegraphics[width=0.49\linewidth]{nDGP1_Opt_hbug.pdf}
 \includegraphics[width=0.49\linewidth]{nDGP2_Opt_hbug.pdf}
 \caption{68.3\% and 95.4\% joint marginal error contours on the cosmological parameters for nDGP1 (left panel) and nDGP2 (right panel) in the optimistic case. In red, we plot \GCsp, in blue, we plot  WL+XC+\GCph\ and in green we plot \GCsp+WL+XC+\GCph.}
 \label{fig:ellipses-nDGP}
\end{figure*}

The two cases considered for the nDGP model are $\log_{10}{\Omega_{\rm rc}}=-0.60206$ (nDGP1) and $\log_{10}{\Omega_{\rm rc}=-6}$ (nDGP2). We summarise the 68.3\%  marginalised errors (relative to the fiducial values) in Table~\ref{tab:rel-errors-DGP} on the cosmological and model parameters in the optimistic, pessimistic and quasi-linear case. We show in Fig.~\ref{fig:ellipses-nDGP} the 68.3\% and 95.4\% joint marginal error contours on cosmological and model parameters for the optimistic scenario for both nDGP1 (left panel) and nDGP2 (right panel). In Fig.~\ref{fig:ellipses-nDGP-pes} we show the same but for the pessimistic setting.

In the optimistic setting, we find that \Euclid will be able to constrain $\log_{10}{\Omega_{\rm rc}}$ as follows:
\begin{itemize}
 \item nDGP1:
 \begin{itemize}
     \item at 93.4\% for \GCsp\ alone;
     \item at 20\% when considering WL+XC+\GCph;
     \item at 15\% for \GCsp+WL+XC+\GCph.
 \end{itemize}
 \item nDGP2: \begin{itemize}
     \item \GCsp\ has no power in constraining the parameter.
     \item at 81\% for both WL+XC+\GCph\ and \GCsp+WL+XC+\GCph.
 \end{itemize}
\end{itemize}  

Regardless of the specific case considered (nDGP1 or nDGP2), the power in constraining $\Omega_{\rm m,0}$ and $\Omega_{\rm b,0}$ is the same, while $n_{\rm s}$ and $\sigma_8$ are better constrained in the nDGP2 case and $h$ in the nDGP1 case. For example, in the optimistic setting, $n_{\rm s}$ for WL+XC+\GCph\ is constrained at $0.52\%$ for nDGP1 and $0.37\%$ for nDGP2 and they further improve when considering the full combination of probes, when they are $0.27\%$ and $0.19\%$, respectively. In Fig.~\ref{fig:ellipses-nDGP} we note in the $\log_{10}{\Omega_{\rm rc}}$--$n_{\rm s}$ panel that the two parameters are anti-correlated in the nDGP1 case. This makes the forecast errors on $n_{\rm s}$ larger than the nDGP2 case where the anti-correlation disappears. A similar discussion can be made for the $\sigma_8$ parameter. In the optimistic setting, for nDGP1, we have $0.37\%$ for WL+XC+\GCph\ and $0.21\%$ for the full combination of probes, while for the nDGP2 model we obtain $0.25\%$ and $0.14\%$, respectively. Instead $h$ is better constrained in nDGP1 being 1.25\% and 1.35\% for nDGP2 using \GCsp.

As previously done for the JBD model, we translated the forecast errors on the parameter $\log_{10}{\Omega_{\rm rc}}$ into the parameter $\Omega_{\rm rc}$ [see Eq.~(\ref{Eq:logXtoX})]. 
For the optimistic scenario, we obtain for nDGP1 $\Omega_{\rm rc}=0.25^{+0.66}_{-0.18}$ for \GCsp, $\Omega_{\rm rc}=0.25^{+0.08}_{-0.06}$ for WL+XC+\GCph\ and $\Omega_{\rm rc}=0.25^{+0.06}_{-0.05}$ for \GCsp+WL+XC+\GCph.
These correspond to relative errors that can reach $264\%$, $32\%$, and $24\%$, respectively. For nDGP2, we obtain an upper bound that is $\Omega_{\rm rc} < 0.07$ in the optimistic setting, while for the pessimistic case, $\Omega_{\rm rc}$ is unconstrained regardless of the probe we consider because the fiducial parameter is very close to \lcdm\ and we do not have enough non-linear information. To examine this, we considered the linear power spectrum: We find that the derivative with respect to $\log_{10} \Omega_{\rm rc}$ approaches 0 in the \lcdm\ limit, and $P_{\rm L}'|_{10^{-6}} \approx 0.0002$  compared to $P_{\rm L}'|_{0.25} \approx 0.08$, where a prime here denotes a derivative with respect to $\log_{10}\Omega_{\rm rc}$. This leads us to conclude that the pessimistic case is still too dominated by linear structure information to constrain $\Omega_{\rm rc}$ when we are close to the \lcdm\ limit. It should also be noted that \cite{Bose:2020wch} reported meaningful constraints using cosmic shear with the same scale cuts we employed in our pessimistic case, but they used a full Markov chain Monte Carlo (MCMC) approach to a \lcdm\ fiducial data vector, sampling in $\Omega_{\rm rc}$, as well as employing a different covariance and binning scheme. Additionally, it is known that the results obtained with a Fisher approach can be different from those computed with the MCMC method even though the sampled parameter is the same \citep{Perotto:2006rj,Wolz:2012sr}. 
Therefore a direct comparison of the results is not straightforward. Lastly, our loss in sensitivity in the \lcdm\ limit justifies the tight upper limit in nDGP1 when compared to nDGP2.

Moreover, a cut in the maximum multipole and scale at lower values, as for the pessimistic (or quasi-linear) scenario, leads to larger uncertainties on the model parameter ($\log_{10}\Omega_{\rm rc}$) compared to the optimistic setting, as shown in Table~\ref{tab:rel-errors-DGP} and in Fig.~\ref{fig:barplot}, where the impact of the different scales is highlighted. Conversely, the relative errors on the cosmological parameters deteriorate only slightly. This means that most of the power of \Euclid in constraining this model comes from the non-linear scales.

Finally, in this analysis we forecast the logarithm of $\Omega_{\rm rc}$ following the motivation discussed in Sect. \ref{sec:results}, but we tested that a similar constraining power can be directly obtained by sampling $\Omega_{\rm rc}$. For the $3x2$pt in the optimistic setting we obtain for nDGP1 a 24\% constraint with respect to the 32\% obtained on $\Omega_{\rm rc}$ indirectly from the log sampling.

\begin{table*}[htbp]
\caption{Forecast $1\sigma$ marginal errors on the cosmological and model parameters relative to their corresponding fiducial value for nDGP1 and nDGP2 in the pessimistic, quasi-linear and optimistic cases, using \Euclid observations of \GCsp, WL+XC+\GCph\ and \GCsp+WL+XC+\GCph.
}
\begin{tabularx}{\textwidth}{Xcccccc}
\hline
\rowcolor{jonquil} \multicolumn{7} {c} {{{\mbox{\textbf{nDGP1}}\,\,\, \boldsymbol{$\Omega_{\rm rc}=0.25$}}}} \\ 
 & \multicolumn{1}{c}{$\Omega_{\rm m,0}$} & \multicolumn{1}{c}{$\Omega_{\rm b,0}$} & \multicolumn{1}{c}{$\log_{10}\Omega_{\rm rc}$} & \multicolumn{1}{c}{$h$} & \multicolumn{1}{c}{$n_{\rm s}$} & \multicolumn{1}{c}{$\sigma_{8}$} \\
\hline
\rowcolor{lavender(web)}\multicolumn{7} {l}{{{Pessimistic setting}}} \\ 
\GCsp\ $(k_{\rm max} = 0.15\,h\,{\rm Mpc}^{-1})$ (quasi-linear)  &	2.14\%	&	4.59\%	& 145.65\% & 3.23\%	& 1.82\% & 1.21\%  \\
\GCsp\ $(k_{\rm max} = 0.25\,h\,{\rm Mpc}^{-1})$ &	1.54\%	&	2.51\%	& 101.27\% & 1.51\%	& 0.99\%  &	0.89\%  \\
WL+XC+\GCph & 1.13\% & 5.64\% &	49.59\%	& 2.65\% & 0.92\% &	1.08\% \\
\GCsp+WL+XC+\GCph &	0.66\%	& 2.18\% & 33.98\% & 1.26\%	& 0.49\%	& 0.42\%  \\
\hline
\rowcolor{lavender(web)}\multicolumn{7} {l}{{{Optimistic setting}}}  \\ 
\GCsp\ $(k_{\rm max} = 0.3\,h\,{\rm Mpc}^{-1})$  &	1.46\%	&	2.24\%	&	93.39\%	&	1.25\%	&	0.87\%	&	0.83\%  \\
WL+XC+\GCph &	0.30\%	&	5.08\%	&	19.59\%	&	2.33\%	&	0.52\%	&	0.37\% \\
\GCsp+WL+XC+\GCph &	0.26\%	&	1.62\%	&	14.78\%	&	0.59\%	&	0.27\%	&	0.21\%  \\
\hline
\hline
\rowcolor{jonquil} \multicolumn{7} {c} {{{\mbox{\textbf{nDGP2}}\,\,\, \boldsymbol{$\Omega_{\rm rc}=10^{-6}$}}}} \\ 
\hline
\rowcolor{lavender(web)}\multicolumn{7} {l}{{{Pessimistic setting}}} \\ 
\GCsp\ $(k_{\rm max} = 0.15\,h\,{\rm Mpc}^{-1})$ (quasi-linear)  &	2.19\%	& 5.01\%	&	3053.42\%	&	3.53\%	&	1.94\%	& 2.53\%  \\
\GCsp\ $(k_{\rm max} = 0.25\,h\,{\rm Mpc}^{-1})$ &	1.56\%	&	2.65\%	&	1969.23\%	&	1.65\%	&	1.01\%	&	1.68\%  \\
WL+XC+\GCph &	0.86\%	&	5.61\%	&	398.64\%	&	2.65\%	&	0.81\%	&	0.58\% \\
\GCsp+WL+XC+\GCph &	0.77\%	& 1.89\% & 379.52\% & 1.13\% & 0.77\%	&	0.39\%  \\
\hline
\rowcolor{lavender(web)}\multicolumn{7} {l}{{{Optimistic setting}}}  \\ 
\GCsp\ $(k_{\rm max} = 0.3\,h\,{\rm Mpc}^{-1})$ &	1.49\%	&	2.24\%	&	1842.43\%	&	1.35\%	&	0.89\%	&	1.54\%  \\
WL+XC+\GCph & 0.29\% & 5.08\% & 81.02\%	& 2.32\% & 0.37\% &	0.25\% \\
\GCsp+WL+XC+\GCph &	0.26\%	&	1.39\%	&	80.77\%	&	0.56\%	&	0.19\%	&	0.14\%  \\
\hline
\hline
\end{tabularx}
\label{tab:rel-errors-DGP}
\end{table*}

\subsection{\texorpdfstring{$k$}{k}-mouflage}\label{Sec:ResultsKM}

\begin{figure*}[ht!]
 \centering
 \includegraphics[width=0.49\linewidth]{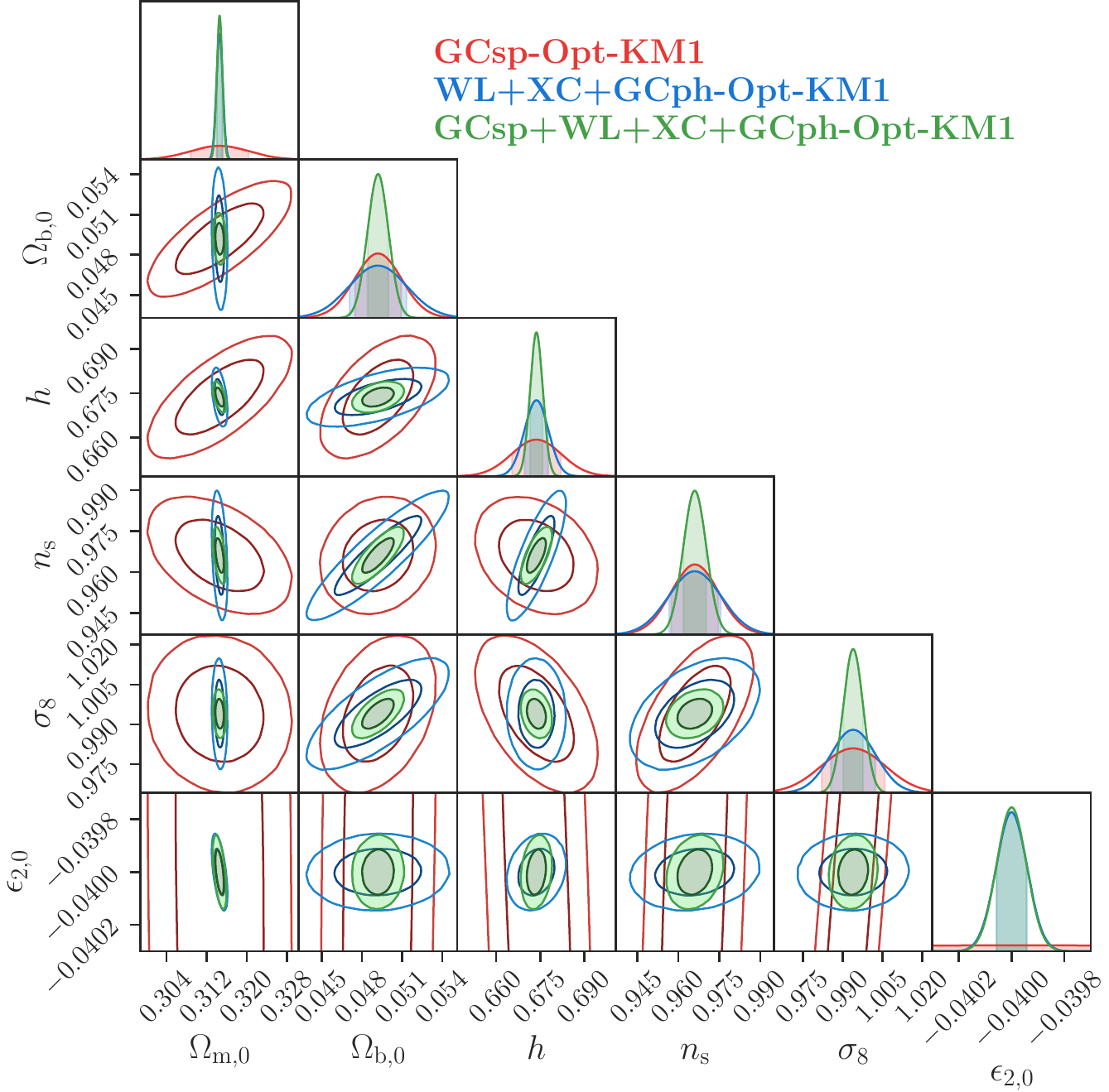}
 \includegraphics[width=0.49\linewidth]{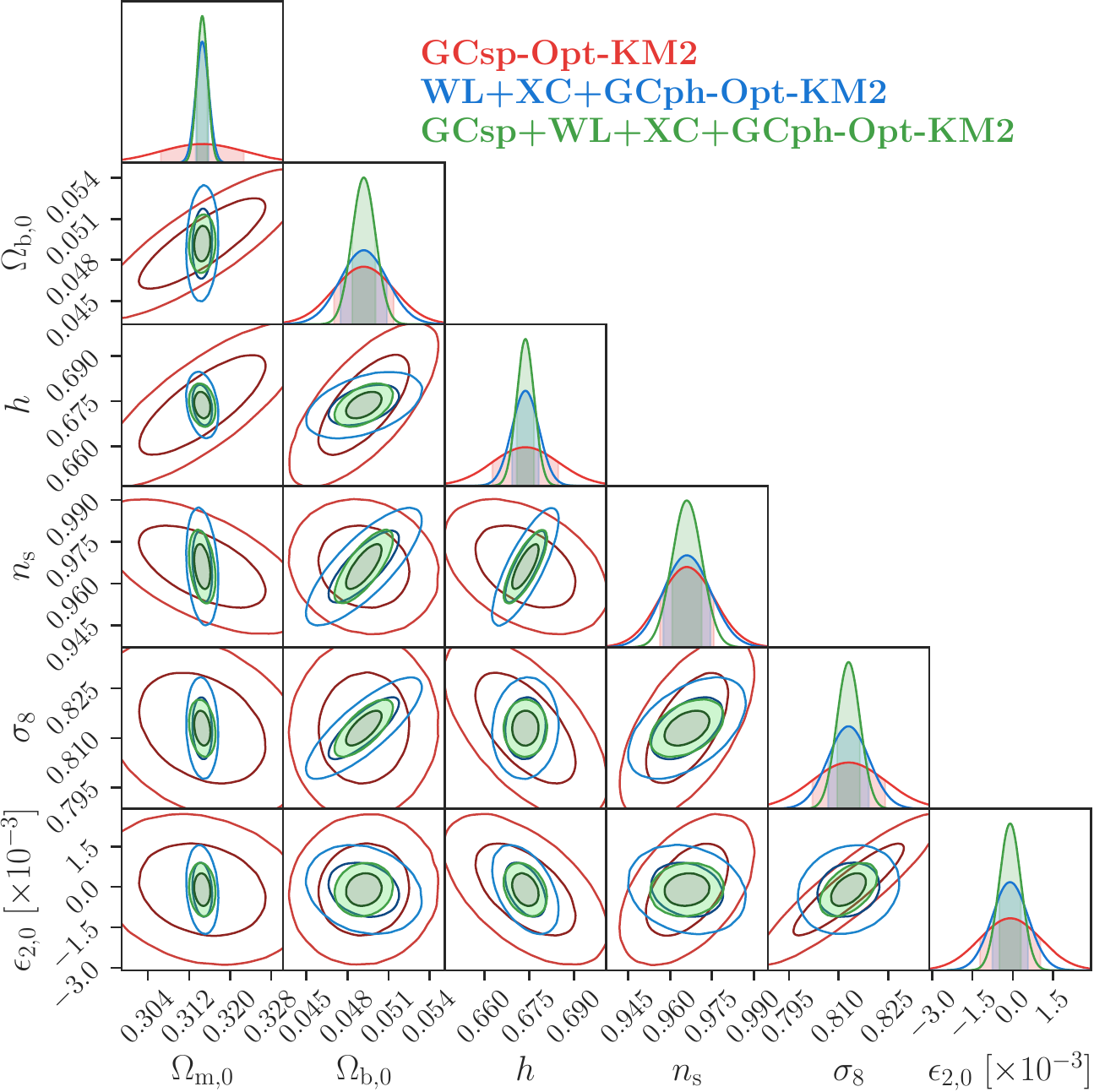}
 \caption{68.3\% and 95.4\% joint marginal error contours on the cosmological parameters for KM1 (left panel) and KM2 (right panel) in the optimistic case. In red, we plot \GCsp, in blue, we plot WL+XC+\GCph\ and in green, we plot \GCsp+WL+XC+\GCph.}
\label{fig:ellipses-KM}
\end{figure*}

We present our results for the KM model with the two fiducial choices of the $\epsilon_{2,0}$ parameter (KM1 $\epsilon_{2,0}= -0.04$ and KM2 $\epsilon_{2,0}= -0.0001$). The KM1 model represents an extreme case that is allowed within the 95.4\% confidence interval by current cosmological data, while KM2 is practically indistinguishable from \lcdm. We show in Fig.~\ref{fig:ellipses-KM} the forecast 68.3\% and 95.4\% joint marginal error contours on cosmological and model parameters for the optimistic scenario. Fig.~\ref{fig:ellipses-KM-pes} shows forecast results for the same models in the pessimistic scenario. The 68.3\% errors for both the optimistic and pessimistic configurations are reported in Table~\ref{tab:rel-errors-KM}.

In the optimistic setting, we find that \Euclid will be able to constrain the modified-gravity parameter $\epsilon_{2,0}$ as follows:
\begin{itemize}
 \item \text{{\rm KM1} }:
 \begin{itemize}
     \item at 3.4\% for the \GCsp\ alone;
     \item at 0.15\% when considering the combination WL+XC+\GCph;
     \item at 0.14\% for \GCsp+WL+XC+\GCph.
 \end{itemize}
 \item \text{{\rm KM2}}: 
 \begin{itemize}
 \item none of the combinations of probes we consider are able to constrain KM2;
 \item upper bound at $10^{-3}$ for the \GCsp\ alone;
 \item upper bound at $7 \times 10^{-4}$ when considering the combination WL+XC+\GCph;
 \item upper bound at $4 \times 10^{-4}$ for \GCsp+WL+XC+\GCph.
\end{itemize}
\end{itemize}

For the KM1 fiducial model, the optimistic configuration will allow us to constrain $\epsilon_{2,0}$ to $\sim 3.4\%$ accuracy using \GCsp\ alone. In the worst-case scenario of cutting the observed galaxy power spectrum at $k_{\rm max} = 0.15\,h\,{\rm Mpc}^{-1}$, the error on $\epsilon_{2,0}$ increases to $\sim 5.5\%$. The constraining power greatly improves for the combination WL+XC+\GCph, and the percentage error on $\epsilon_{2,0}$ decreases to $0.15\%$ in the optimistic case, without a substantial worsening in the pessimistic scenario. Of the cosmological parameters we varied for the KM1 model, $\epsilon_{2,0}$ is most tightly constrained by the combination WL+XC+\GCph. This shows that this combination of data is well suited to capturing the modification induced by $k$-mouflage to the Poisson and lensing equations, probing $\mu$ and $\Sigma$ independently. Moreover, the \Euclid survey will probe the redshift range $z \gtrsim 1$, where the largest effect determined by the running of Newton’s constant is expected to manifest \citep[see, e.g.,][]{Benevento_2019}. 
The fact that with the photometric probes we obtain the tightest relative constraints for the KM1 model, compared to the other two models studied in this work, is consistent with the observation of the lower panels of Figure \ref{fig:MG-observables}, which show that the lensing angular power spectrum is more distant to its \lcdm{} counterpart than in the other two models.
We note that the addition of \GCsp\ to WL+XC+\GCph\ does not improve the constraints on $\epsilon_{2,0}$ in either the pessimistic and optimistic scenarios.

The percentage error on $\epsilon_{2,0}$ in the \lcdm-proximate KM2 fiducial model decreases from $1120\%$ using \GCsp\ alone to a value of $397\%$ with the full combination of \GCsp+WL+XC+\GCph in the best-case scenario. In the optimistic case, the \Euclid survey will therefore be able to detect a deviation from \lcdm\ only when $\vert \epsilon_{2,0} \vert \gtrsim 4 \times 10^{-4}$, which improves over the constraining power of present CMB and LSS data by somewhat more than one order of magnitude.

Fig.~\ref{fig:ellipses-KM} shows that $\epsilon_{2,0}$ is anti-correlated with the $h$ parameter. This correlation can be exploited to reduce the Hubble tension, as noted in \cite{Benevento_2019}. This effect determines a lower forecast error for the $h$ parameter in the KM1 fiducial model, for which $\epsilon_{2,0}$ is tightly constrained by \Euclid probes, especially when the full data combination is considered. A similar argument applies to the $\Omega_{\rm m,0}$ parameter, which is also better constrained for KM1. The uncertainties for the other parameters do not vary much with the fiducial choice of $\epsilon_{2,0}$.


\begin{table*}[htbp]
\caption{Forecast $1\sigma$ marginal errors on the cosmological and model parameters, relative to their corresponding fiducial value, for KM1 and KM2 in the pessimistic, quasi-linear and optimistic cases, using \Euclid observations of \GCsp,
WL+XC+\GCph\ and \GCsp+WL+XC+\GCph.
}
\begin{tabularx}{\textwidth}{Xcccccc}
\hline
\rowcolor{jonquil} \multicolumn{7} {c} {{{\mbox{\textbf{KM1}}\,\,\, \boldsymbol{$\epsilon_{2,0}=-0.04$}}}} \\ 
 & \multicolumn{1}{c}{$\Omega_{\rm m,0}$} & \multicolumn{1}{c}{$\Omega_{\rm b,0}$} & \multicolumn{1}{c}{$\epsilon_{2,0}$} & \multicolumn{1}{c}{$h$} & \multicolumn{1}{c}{$n_{\rm s}$} & \multicolumn{1}{c}{$\sigma_{8}$} \\
\hline
\rowcolor{lavender(web)}\multicolumn{7} {l}{{{Pessimistic setting}}} \\ 
\GCsp\ $(k_{\rm max} = 0.15\,h\,{\rm Mpc}^{-1})$ (quasi-linear) &	6.53\%	&	10.43\%	&	5.48\%	&	3.88\%	&	1.95\%	&	2.39\%  \\
\GCsp\ $(k_{\rm max} = 0.25\,h\,{\rm Mpc}^{-1})$ &	2.23\%	&	4.02\%	&	3.75\%	&	1.52\%	&	1.10\%	&	1.31\%  \\
WL+XC+\GCph &	0.36\%	&	7.36\%	&	0.22\%	&	1.40\%	&	1.93\%	&	1.10\% \\
\GCsp+WL+XC+\GCph &	0.31\%	&	1.77\%	&	0.22\%	&	0.44\%	&	0.63\%	&	0.46\%  \\
\hline
\rowcolor{lavender(web)}\multicolumn{7} {l}{{{Optimistic setting}}}  \\ 
\GCsp\ $(k_{\rm max} = 0.3\,h\,{\rm Mpc}^{-1})$  &	1.82\%	&	3.50\%	&	3.40\%	&	1.23\%	&	0.89\%	&	1.19\%  \\
WL+XC+\GCph &	0.20\%	&	4.30\%	&	0.15\%	&	0.59\%	&	0.98\%	&	0.84\% \\
\GCsp+WL+XC+\GCph &	0.17\%	&	1.57\%	&	0.14\%	&	0.32\%	&	0.43\%	&	0.37\%  \\
\hline
\hline
\rowcolor{jonquil} \multicolumn{7} {c} {{{\mbox{\textbf{KM2}}\,\,\, \boldsymbol{$\epsilon_{2,0}=-0.0001$}}}} \\ 
\hline
\rowcolor{lavender(web)}\multicolumn{7} {l}{{{Pessimistic setting}}} \\ 
\GCsp\ $(k_{\rm max} = 0.15\,h\,{\rm Mpc}^{-1})$ (quasi-linear) &	7.85\%	&	12.28\%	&	1720.00\%	&	4.48\%	& 2.24\% &	2.84\%  \\
\GCsp\ $(k_{\rm max} = 0.25\,h\,{\rm Mpc}^{-1})$ &	3.19\%	&	5.41\%	&	1190.00\%	&	1.99\%	&	1.13\%	&	1.47\%  \\
WL+XC+\GCph &	1.18\%	&	5.22\%	&	939.23\%	& 1.03\%	&	1.51\%	&	1.17\% \\
\GCsp+WL+XC+\GCph &	0.73\%	&	2.23\%	&	642.65\%	&	0.79\%	&	0.82\%	&	0.80\%  \\
\hline
\rowcolor{lavender(web)}\multicolumn{7} {l}{{{Optimistic setting}}}  \\ 
\GCsp\ $(k_{\rm max} = 0.3\,h\,{\rm Mpc}^{-1})$  &	2.57\%	&	4.41\%	&	1120.00\%	&	0.62\%	&	1.00\%	&	1.34\%  \\
WL+XC+\GCph &	0.40\%	&	3.43\%	&	658.67\%	&	0.66\%	&	0.87\%	&	0.75\% \\
\GCsp+WL+XC+\GCph &	0.33\%	&	1.73\%	&	396.60\%	&	0.43\%	&	0.55\%	&	0.42\%  \\
\hline
\hline
\end{tabularx}
\label{tab:rel-errors-KM}
\end{table*}

\section{Conclusions} \label{sec:conclusions}

We have explored the constraining power of the future \Euclid mission for linearly scale-independent extensions of the concordance cosmological model, that is, models that induce modifications to the linear growth of perturbations that are solely time-dependent while featuring different testable types of screening mechanisms at smaller non-linear scales. We considered three specific models, namely JBD, a scalar-tensor theory with a standard kinetic term and a flat potential (Sect.~\ref{Sec:theoryJBD}); the nDGP gravity, a braneworld model in which our Universe is a four-dimensional brane embedded in a five-dimensional Minkowski space-time (Sect.~\ref{Sec:theorynDGP}); and KM gravity, an extension of $k$-essence scenarios with a universal coupling of the scalar field to matter (Sect.~\ref{Sec:theoryKM}).

We derived forecasts from \Euclid spectroscopic and photometric primary probes on the cosmological parameters and the additional parameters of the models, $\log_{10}\omega_{\rm BD}$ for JBD, $\log_{10}\Omega_{\rm rc}$ for nDGP, and $\epsilon_2$ for KM. In order to do this, we applied the Fisher matrix method to weak lensing (WL), photometric galaxy clustering (\GCph), spectroscopic galaxy clustering (\GCsp), and the cross-correlation (XC) between \GCph\ and WL. For each MG model, we considered two fiducial values for the corresponding model parameter, following the rationale of a case that is representative of the \lcdm\ limit and another that differs more significantly while still being (nearly) compatible with current bounds.  
We modelled the non-linear matter power spectrum using different prescriptions for each MG model: The \texttt{HMCODE} \citep{Mead:2015yca,Mead:2016zqy} calibrated on a suite of {\it N}-body simulations from modified versions of COLA \citep{Tassev:2013pn,Winther:2017jof} and \texttt{RAMSES} \citep{Teyssier:2001cp} for JBD; the halo model reaction \citep{Cataneo:2018cic} for nDGP; and an analytical approach that combined one-loop perturbation theory with a halo model for KM following \cite{Brax:2014yla,Brax:2015lra}. 

When setting the \Euclid survey specifications, we defined three scenarios that were characterized by different cuts in the maximum multipole and wavenumber to assess the constraining power of non-linear scales: the quasi-linear scenario with $k_{\rm max}=0.15\,h\,{\rm Mpc}^{-1}$ for \GCsp; the pessimistic scenario with $k_{\rm max}=0.25\,h\,{\rm Mpc}^{-1}$ for \GCsp, $\ell_{\rm max}=1500$ for WL, and $\ell_{\rm max}=750$ for \GCph; and the optimistic scenario with $k_{\rm max}=0.3\,h\,{\rm Mpc}^{-1}$ for \GCsp, $\ell_{\rm max}=5000$ for WL, and $\ell_{\rm max}=3000$ for \GCph.
In Sect.~\ref{sec:results} we discussed and reported in great detail the results corresponding to these different scenarios. For each case, we considered three different combinations of data: \GCsp\ alone, WL+XC+\GCph\ and the full \GCsp+WL+XC+\GCph. 
For each of these combinations we presented the relative errors on the fiducial values of the three models as they can provide indications about possible future detections of non-zero modified-gravity parameters.

With the full set of probes, we found that in the optimistic scenario, \Euclid alone will be able to constrain the JBD parameter $\omega_{\rm BD}=800^{+200}_{-160}$ in the fiducial case JBD1 for the full combination of probes, and in the JBD2 case $\omega_{\rm BD}=2500^{+1070}_{-750}$. This indicates that \Euclid alone will be capable of detecting at a statistically significant level only values of $\omega_{\rm BD}$ that are compatible with the current bounds from CMB and LSS.

For the nDGP model, we obtain for the optimistic scenario that $\Omega_{\rm rc}=0.25^{+0.06}_{-0.05}$ with the full combination in the nDGP1 fiducial case, while for the nDGP2 fiducial ($\Omega_{\rm rc} = 10^{-6}$) we find that $\Omega_{\rm rc} < 0.07$. 
Furthermore, the nDGP2 case is closer to the \lcdm\ model than the KM2 model. The constraints of nDGP2 are therefore worse than those of KM2.

For the KM scenario, we obtain constraints for the full combination of probes in the fiducial KM1 case of $\epsilon_{2,0}= -0.04\pm 5.6\times 10^{-5}$. Out of the full set of cosmological plus model parameters, $\epsilon_{2,0}$ is the most tightly constrained parameter for KM1 by the full combination because of the independent constraints on clustering and lensing, which probe the running of Newton's constant in the range $z \gtrsim 1$ where the strongest effect is expected. The KM2 fiducial model (with $\epsilon_{2,0}= -0.0001$) is unconstrained for all the different combinations of probes.

Although it has been shown that for values close to the $\Lambda$CDM limit, the nDGP2 and KM2 models remain unconstrained in the pessimistic setting, it is remarkable that \Euclid alone can significantly reduce the allowed space for these models. For the full probe combination (\GCsp+WL+XC+\GCph) the nDGP2 upper bound at $\Omega_{\rm rc}<0.07$ for the optimistic scenario significantly improves the current constraints derived with the same approach and similar type of data. This is consistent with the results found in \cite{Bose:2020wch}.

The forecasts for KM2 give $\epsilon_{2,0}|<4\times 10^{-4}$ in the optimistic case and $|\epsilon_{2,0}|<2\times 10^{-3}$ in the pessimistic case. This improves the current constraints by more than one order of magnitude, even in the pessimistic case, by combining different cosmological datasets \citep{Benevento_2019}. In addition, \Euclid alone will provide constraints on $\omega_{\rm BD}$ that will be tighter than those obtained by a combination of current CMB and low-redshift data \cite{Joudaki:2020shz}.

Finally, our analysis clearly showed that most of the constraining power of \Euclid comes from the non-linear scales. We conclude that \Euclid will be able to provide outstanding constraints on extensions beyond the concordance model given a good modelling of our theoretical observables at these scales, such as we used in this analysis, seen from the substantial differences in constraining power between the optimistic and pessimistic case. The current dedicated modelling of non-linearities for spectroscopic galaxy clustering, developed and tested for \lcdm\ and applied to BOSS data by \cite{DAmico:2019fhj,Ivanov:2019pdj,Chen:2021wdi}, for instance, can be straightforwardly extended to the class of scale-independent models considered in this paper, as was already shown, in the nDGP model in \cite{Piga:2022mge}, for example. 
Additionally we would like to stress that the combination and cross-correlation of future \Euclid data with CMB measurements will be of crucial importance for the cosmological constraints of extended models. As already shown within the \Euclid Collaboration \citep{Euclid:2021qvm}, the improvements in the constraints when \Euclid is cross-correlated with CMB data can be of the order of two to three and in some cases even larger. A large effort is currently ongoing to extend this analysis to other extended models, including those in this work.



\begin{acknowledgements}
\AckEC
B.B. and L.L. were supported by a Swiss National Science Foundation Professorship grant (Nos. 170547 \& 202671).
E.B.~has received funding from the European Union’s Horizon 2020 research and innovation programme under the Marie Sk\l{}odowska-Curie grant agreement No 754496.
B.B was supported by a UK Research and Innovation Stephen Hawking Fellowship (EP/W005654/2). 
P.G.F. acknowledges funding from STFC, the Beecroft Trust and the European Research Council (ERC) under the European Unions Horizon 2020 research and innovation programme (grant agreement No 693024).
N.F. is supported by the Italian Ministry of University and Research (MUR) through Rita Levi Montalcini project  ``Tests of gravity at cosmological scales" with reference PGR19ILFGP. N.F., F.P. and A.R.F. also acknowledge the FCT project with ref. number PTDC/FIS-AST/0054/2021.
K.K. is supported by the UK STFC grant ST/S000550/1 and ST/W001225/1. For the purpose of open access, the author(s) has applied a Creative Commons Attribution (CC BY) licence to any Author Accepted Manuscript version arising.
F.P. acknowledges partial support from the INFN grant InDark and the Departments of Excellence grant L.232/2016 of the Italian Ministry of University and Research (MUR).
A.R.F. acknowledges support from DL 57/2016 from the ‘Departamento de F\'isica, Faculdade de Ci\^encias, Universidade de Lisboa’. 
I.T. acknowledges funding from the European Research Council (ERC) under the European Union's Horizon 2020 research and innovation programme (Grant agreement No.\ 863929; project title ``Testing the law of gravity with novel large-scale structure observables'' and acknowledges support from the Spanish Ministry of Science, Innovation and Universities through grant ESP2017-89838, and the H2020 programme of the European Commission through grant 776247. Z.S. acknowledges funding from DFG project 456622116. G.B. acknowledges support from ASI/INFN grant n. 202one-43-HH.0. MP acknowledges support by the MIUR `Progetti di Ricerca di Rilevante Interesse Nazionale' (PRIN) Bando 2022 - grant 20228RMX4A.

\end{acknowledgements}


\bibliographystyle{aa}
\bibliography{biblio}

\newpage
\appendix

\section{Further results} \label{sec:appendix}

\begin{figure*}[t!]
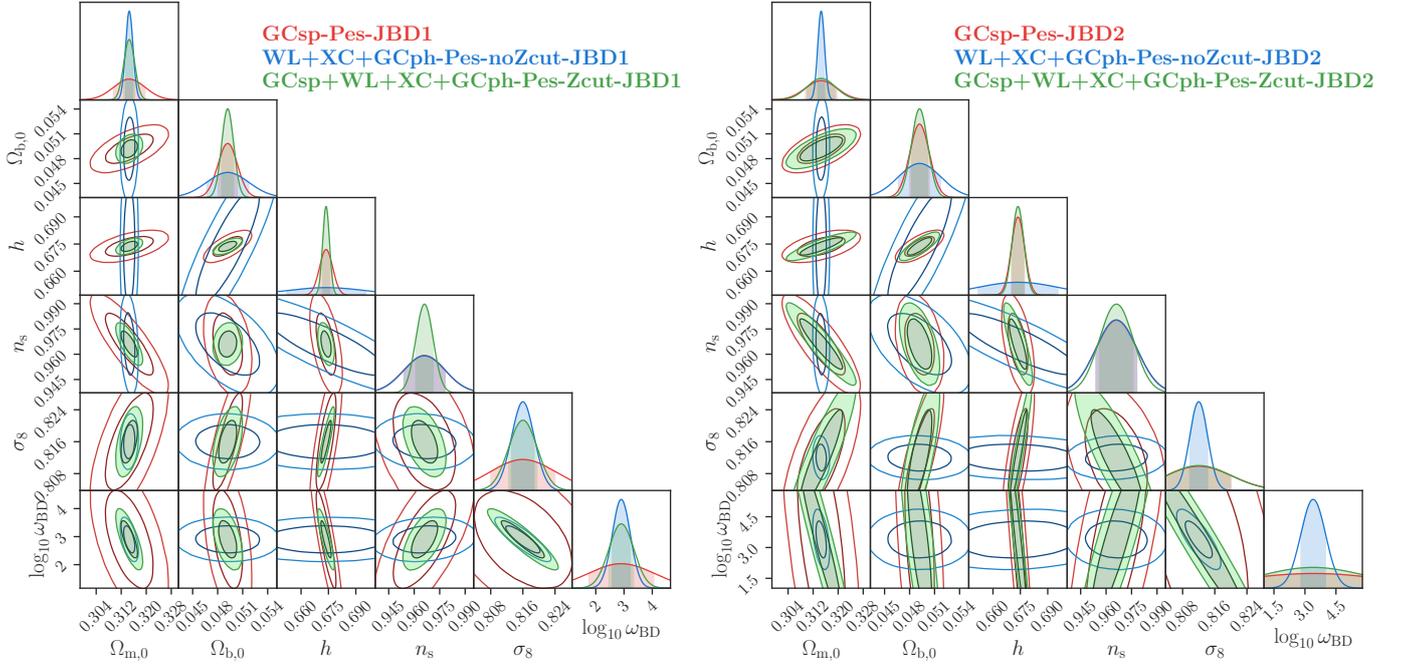

 \centering
 \includegraphics[width=0.49\linewidth]{JBD1_Pes_hbug.pdf}
 \includegraphics[width=0.49\linewidth]{JBD2_Pes_hbug.pdf}
 \caption{68.3\% and 95.4\%  joint marginal error contours on the cosmological parameters for JBD1 (left panel) and JBD2 (right panel) in the pessimistic case. In red, we plot \GCsp, in blue, we plot  WL+XC+\GCph\ and in green, we plot \GCsp+WL+XC+\GCph.}
\label{fig:ellipses-JBD-pes}
\end{figure*}

\begin{figure*}[t!]
 \centering
 \includegraphics[width=0.49\linewidth]{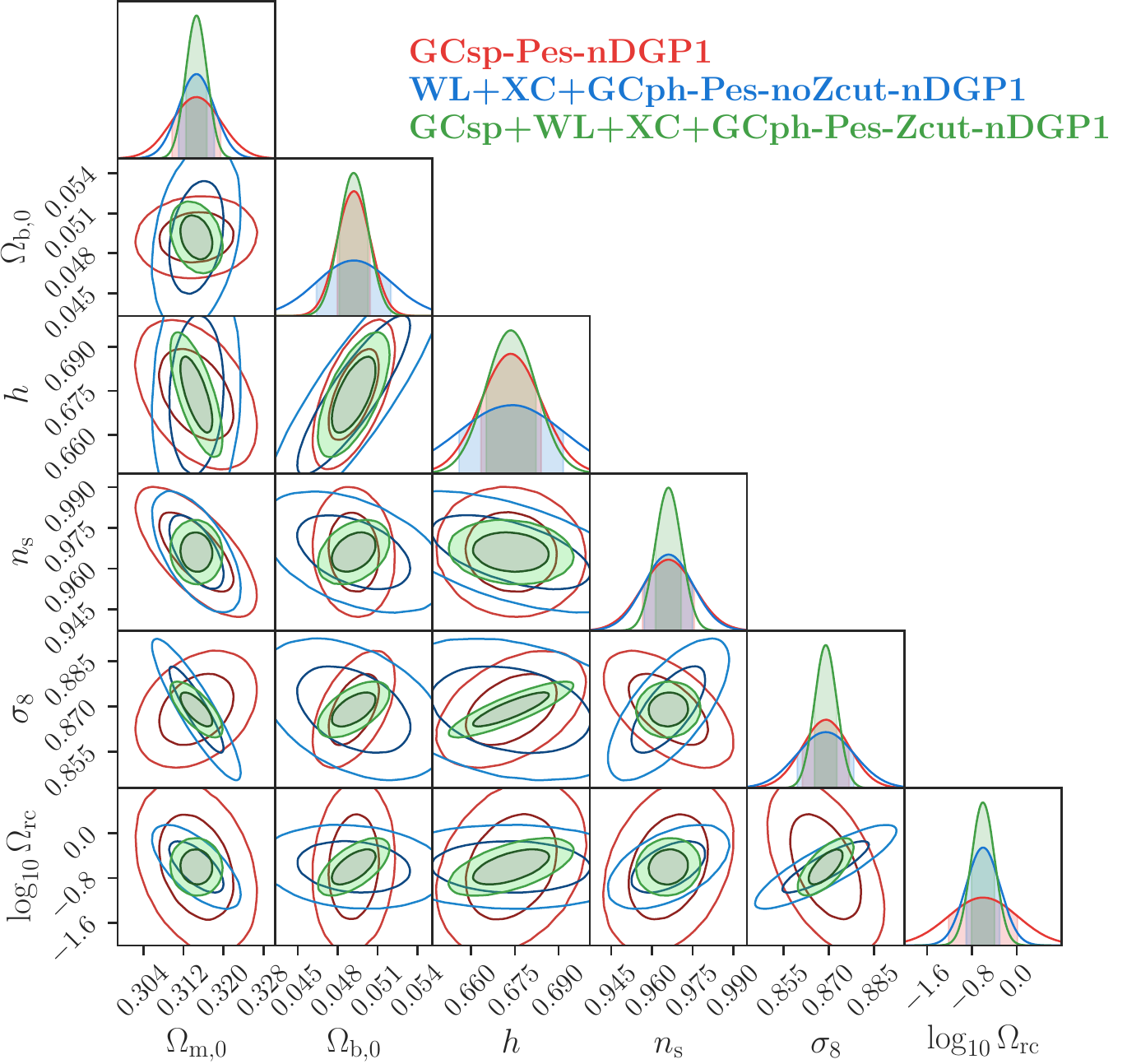}
 \includegraphics[width=0.49\linewidth]{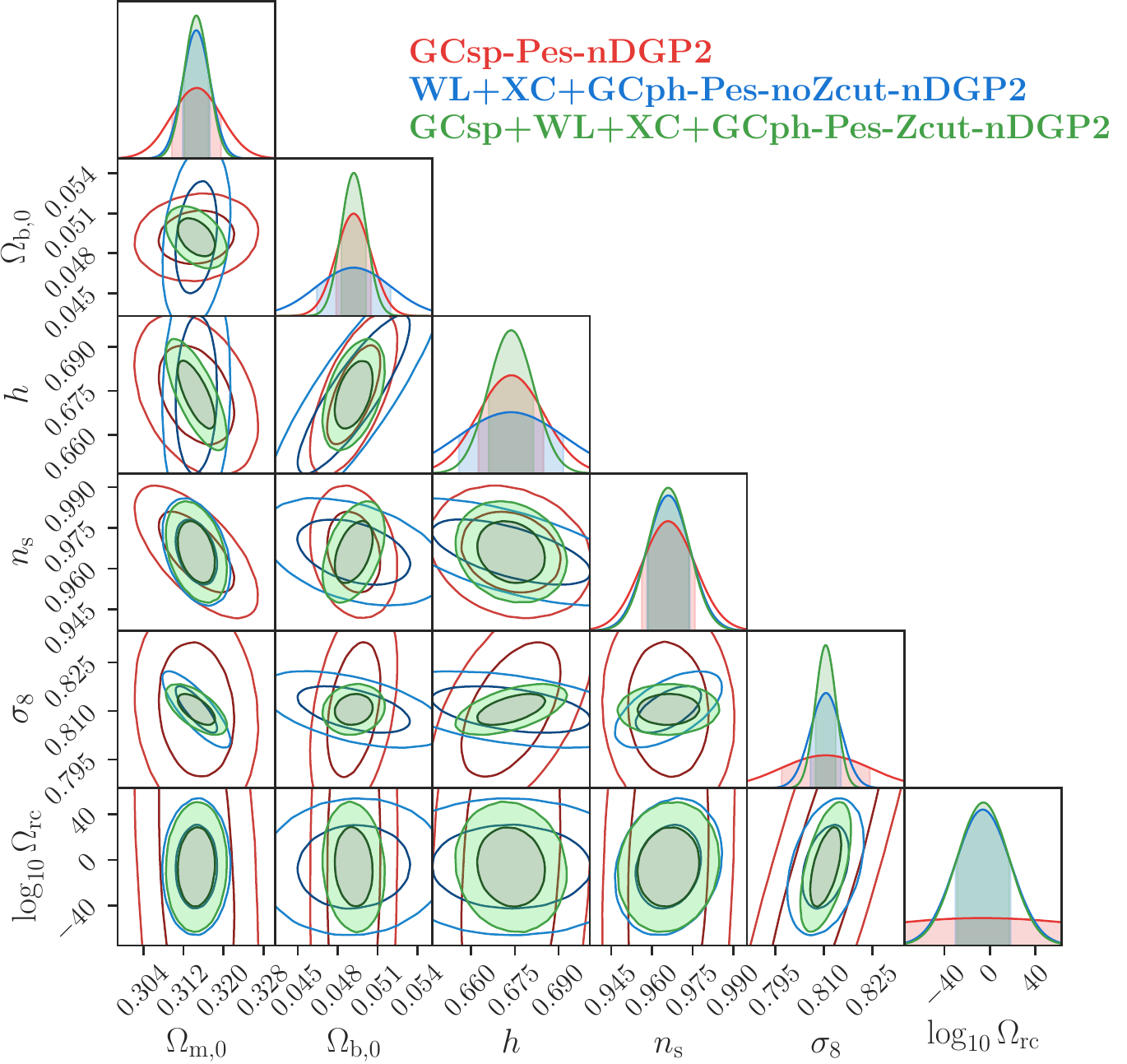}
 \caption{68.3\% and 95.4\% joint marginal error contours on the cosmological parameters for nDGP1 (left panel) and nDGP2 (right panel) in the pessimistic case. In red, we plot \GCsp, in blue, we plot  WL+XC+\GCph\ and in green, we plot \GCsp+WL+XC+\GCph.}
 \label{fig:ellipses-nDGP-pes}
\end{figure*}

\begin{figure*}[t!]
 \centering
 \includegraphics[width=0.49\linewidth]{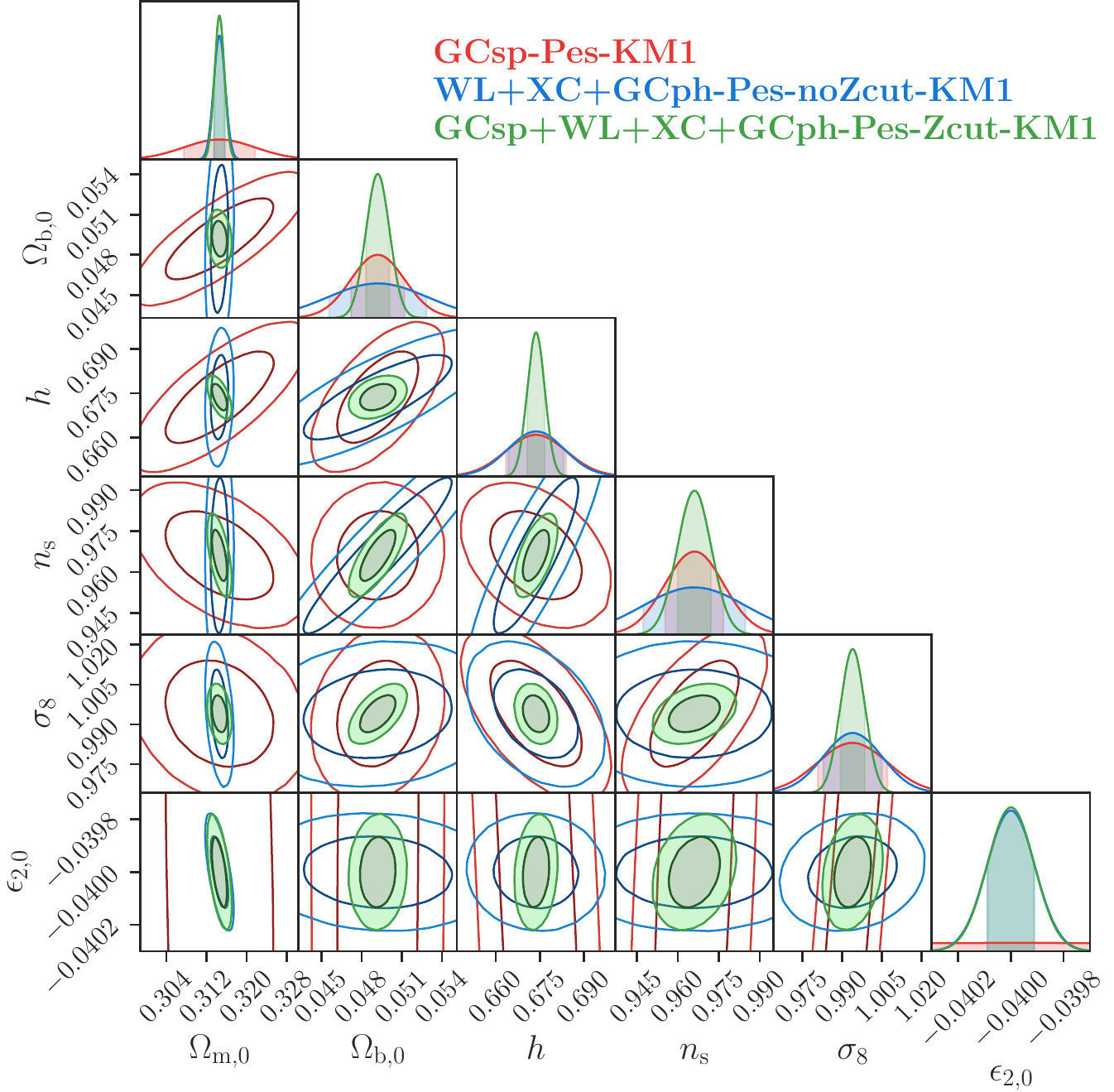}
 \includegraphics[width=0.49\linewidth]{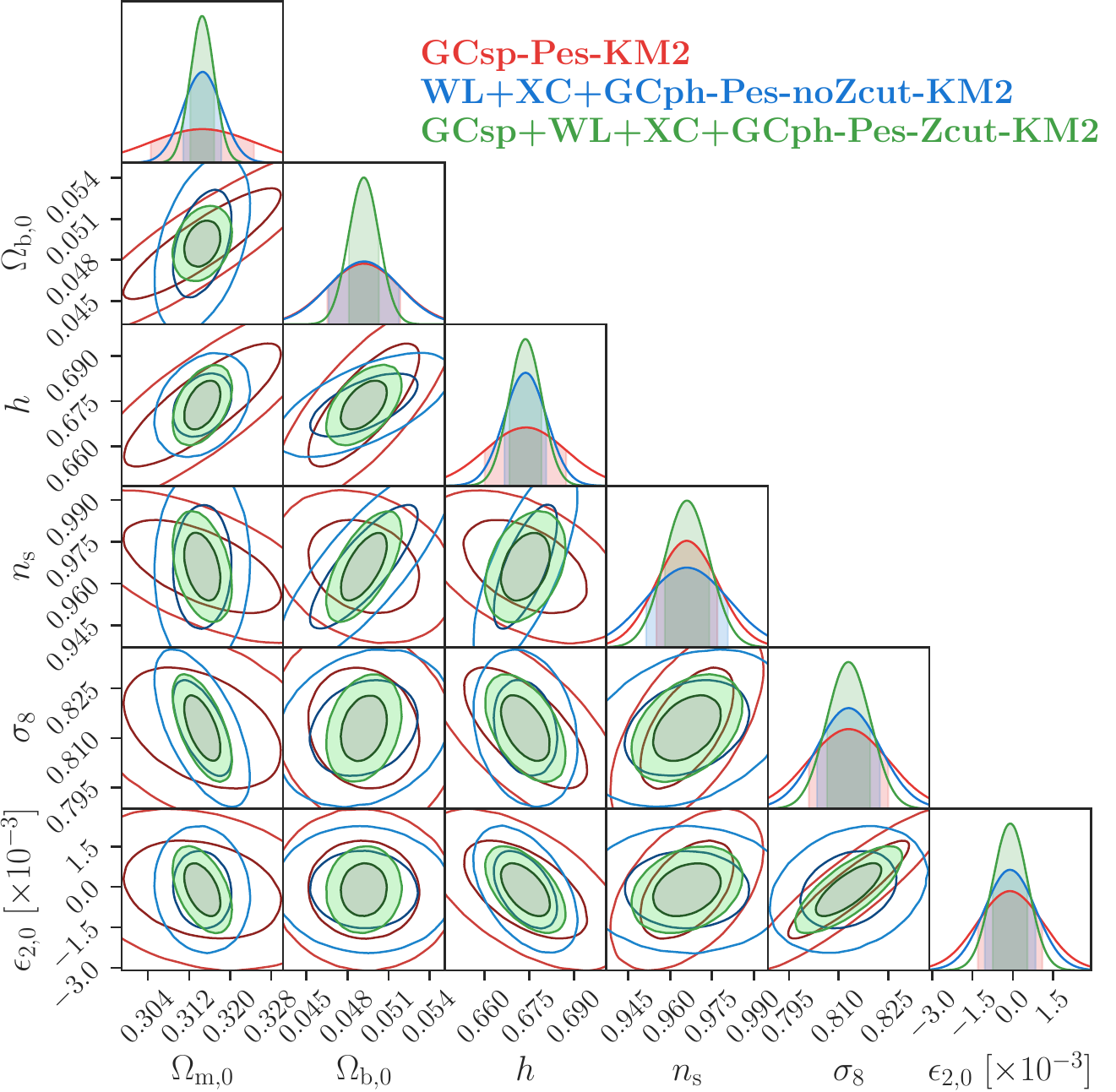}
 \caption{68.3\% and 95.4\%  joint marginal error contours on the cosmological parameters for KM1 (left panel) and KM2 (right panel) in the pessimistic case. In red, we plot \GCsp, in blue, we plot  WL+XC+\GCph\ and in green, we plot \GCsp+WL+XC+\GCph.}
 \label{fig:ellipses-KM-pes}
\end{figure*}

In this appendix we show additional results to complement what we discussed in Sect.~\ref{sec:results}. In Figs.~\ref{fig:ellipses-JBD-pes}, \ref{fig:ellipses-nDGP-pes}, and \ref{fig:ellipses-KM-pes} we provide the 68.3\% and 95.4\% joint marginal error contours in the pessimistic case on the cosmological parameters for JBD, nDGP, and KM respectively.

\end{document}